\documentclass[onecolumn,showpacs,preprintnumbers]{revtex4}
\usepackage{mathrsfs}
\usepackage{graphicx}
\usepackage{dcolumn}
\usepackage{bm}
\usepackage{epsfig}
\usepackage{amsmath,amssymb}
\usepackage{multirow}
\usepackage{placeins}
\usepackage{caption}
\usepackage{array}
\usepackage{float}
\usepackage{subfigure}
\usepackage{pdflscape}
\usepackage[figuresleft]{rotating}

\newcommand{\PreserveBackslash}[1]{\let\temp=\\#1\let\\=\temp}

\newcolumntype{C}[1]{>{\PreserveBackslash\centering}p{#1}}

\newcolumntype{R}[1]{>{\PreserveBackslash\raggedleft}p{#1}}

\newcolumntype{L}[1]{>{\PreserveBackslash\raggedright}p{#1}}

\begin{document}
\title{The $S$- and $P$-wave fully charmed tetraquark states and their radial excitations}
\author{Guo-Liang Yu$^{1}$}
\email{yuguoliang2011@163.com}
\author{Zhen-Yu Li$^{2}$}
\email{zhenyvli@163.com}
\author{Zhi-Gang Wang$^{1}$}
\email{zgwang@aliyun.com}
\author{Lu Jie$^{1}$}
\author{Yan Meng$^{1}$}

\affiliation{$^1$ Department of Mathematics and Physics, North China
Electric Power University, Baoding 071003, People's Republic of
China\\$^2$ School of Physics and Electronic Science, Guizhou Education University, Guiyang 550018, People's Republic of
China}
\date{\today }

\begin{abstract}
Inspired by recent progresses in observations of the fully charmed tetraquark states by LHCb, CMS, and ATLAS collaborations, we perform a systematic study of the ground states and the first radial excitations of the $S$- and $P$-wave $\mathrm{cc}\bar{\mathrm{c}}\bar{\mathrm{c}}$ system. Their mass spectra, root mean square(r.m.s.) radii and radial density distributions are studied with the relativized quark model. The calculations show that there is no stable bound states for the fully charmed tetraquark states, and the r.m.s. radii of these tetraquark states are smaller than 1 fm. Our results support assigning X(6600) structure, M$_{X(6600)}=6552\pm10\pm12$ MeV, as one of the $0^{++}$(1$S$) and $2^{++}$(1$S$) states or their mixtures. Another structure also named as X(6600) by CMS Collaboration, M$_{X(6600)}=6.62\pm0.03^{+0.02}_{-0.01}$ GeV, may arise from the lowest 1$P$ states with $J^{PC}$=$0^{-+}$, $1^{-+}$, and $2^{-+}$. The possible assignments for X(6900) include the $0^{++}$(2$S$), $2^{++}$(2$S$) states, and the highest 1$P$ state with $J^{PC}=0^{-+}$. As for X(7200), it can be interpreted as one of the highest 2$P$ states with $J^{PC}=0^{-+}$, $1^{-+}$, and $2^{-+}$, and the 3$S$ states can not be completely excluded from the candidates.
\end{abstract}


\maketitle

\begin{Large}
\textbf{1 Introduction}
\end{Large}

Since the first observation of X(3872) by Belle in 2003\cite{3872}, a series of exotic hadrons named as XYZ states emerged like bamboo shoots after a spring rain. These discoveries in experiments have motivated theorists to devote a great deal of energy in studying the inner structure of these new states. In general, these new observed states can not be categorized as the conventional mesons or baryons, and they are commonly explained as compact tetraquark states, hadronic molecular states or admixture of these two states. Among these new hadrons, the states composed of fully heavy quark components are especially interesting. It was supposed that the interactions between heavy quarks(antiquarks) are dominated by the short range one-gluon-exchange(OGE) potential rather than the long-range potential resulted from light meson exchanges. Thus, the configurations comprised of four heavy quarks are easier to form compact tetraquark states instead of the molecules.

In 2020, the LHCb collaboration studied the $J/\psi J/\psi$ invariant mass spectrum using the $pp$ collision data at center-of-mass energies of $\sqrt{s}$=7, 8 and 13 TeV\cite{LHCb6900}. They observed a broad structure above the $J/\psi J/\psi$ threshold ranging from 6.2 to 6.8 GeV and a narrow structure around 6.9 GeV$/c^{2}$ with the significance of larger than 5$\sigma$. At the ICHEP 2022 conference, the ATLAS collaboration reported their observation of several fully charmed tetraquark excesses decaying into a pair of charmonium states in the four $\mu$ final states\cite{ATLAS}. They not only confirmed the existence of the X(6900) structure reported earlier by LHCb Collaboration but also observed a broad structure at lower mass and a structure labelled as X(6600). In addition, the CMS Collaboration also reported their studies on the $J/\psi J/\psi$ mass spectrum, which were carried out by using proton-proton data at center-of-mass energies of 13 TeV\cite{CMS}. The existence of X(6900) was confirmed with significance larger than 9.4$\sigma$. Besides X(6900), the CMS collaboration also observed two structures in the $J/\psi J/\psi$ channel, which were labeled as X(6600) and X(7200). All of the experimental data about the fully charmed tetraquark states are collected in Table I.
\begin{center}
\begin{table}[htp]
\begin{ruledtabular}\caption{The experimental data for the fully charmed tetraquark states.}
\begin{tabular}{c c c c}
 &States & Mass & Width \\ \hline
LHCb model I\cite{LHCb6900}& \multirow{2}{*}{X(6900) } & $6905\pm11\pm7$ MeV & $80\pm19\pm33$ MeV  \\
LHCb model II\cite{LHCb6900} &        &    $6886\pm11\pm1$ MeV     &    $168\pm33\pm69$ MeV    \\ \hline
\multirow{3}{*}{ATLAS\cite{ATLAS} } & X(6200)& $6.22\pm0.03^{+0.04}_{-0.05}$ GeV & $0.31\pm0.12^{+0.07}_{-0.08}$ GeV \\
\multirow{3}{*}{ } & X(6600)& $6.62\pm0.03^{+0.02}_{-0.01}$ GeV &$0.31\pm0.09^{+0.06}_{-0.11}$ GeV\\
\multirow{3}{*}{ } & X(6900)& $6.87\pm0.03^{+0.06}_{-0.01}$ GeV& $0.12\pm0.04^{+0.03}_{-0.01}$ GeV\\ \hline
\multirow{3}{*}{CMS\cite{CMS} } & X(6600)& $6552\pm10\pm12$ MeV & $124\pm29\pm34$ MeV \\
\multirow{3}{*}{ } & X(6900)& $6927\pm9\pm5$ MeV & $122\pm22\pm19$ MeV \\
\multirow{3}{*}{ } & X(7200)& $7287\pm19\pm5$ MeV& $95\pm46\pm20$ MeV\\
\end{tabular}
\end{ruledtabular}
\end{table}
\end{center}

Actually, the full-heavy tetraquark states were already studied in the literatures\cite{Tc1,Tc2,Tc3,Tc4,Tc5,Tc6,Tc7,Tc8,Tc9,Tc10,Tc11,Tc12,Tc13,Tc14,Tc15,Tc16,Tc17,Tc18,Tc19,Tc20,Tc21,Tc22,Tc23,Tc24,Tc25,Tc26,Tc27,Tc28,Tc29,Tc30,Tc31,Tc32} before the experimental observations. The recent breakthroughs in experiments in searching for the fully charmed tetraquark states have inspired again the intensive discussions. People employed many methods/models to carry out their studies on the structure, the production, and decay property of these new states\cite{Tc34,Tc35,Tc36,Tc37,Tc38,Tc39,Tc40,Tc41,Tc42,Tc420,Tc421,Tc422,Tc423,Tc424,Tc425,Tc426,Tc427,Tc428,Tc429,Tc4210,Tc4211,Tc4212,Tc4213,Tc4214,Tc4215,Tc4216,Tc4217,Tc4218,Tc4219,Tc4220,Tc4221,Tc4222,Tc4223}. The most popular interpretations about these new discoveries are compact tetraquark states\cite{Tc43,Tc44,Tc45,Tc46,Tc47,Tc48,Tc49,Tc50,Tc51,Tc52}. In Ref.\cite{Tc30}, Q. F. L\"{u} $et$ $al$. analyzed the mass spectrum of the $S$-wave full-heavy tetraquark states with a extended relativistic quark model. In their studies, the X(6900) was categorized as the first radial excitation of $\mathrm{cc}\bar{\mathrm{c}}\bar{\mathrm{c}}$ system. Using the method of QCD sum rules, Z. G. Wang made possible assignments of the X(6600), X(6900) and X(7300) in the picture of tetraquark states with the J$^{PC}$=0$^{++}$ or 1$^{+-}$\cite{Tc12,Tc13,Tc14,Tc34}. In Ref.\cite{Tc36}, G. J. Wang $et$ $al$. studied the mass spectra of the $S$- and $P$-wave $\mathrm{cc}\bar{\mathrm{c}}\bar{\mathrm{c}}$ and $\mathrm{bb}\bar{\mathrm{b}}\bar{\mathrm{b}}$ systems with a nonrelativistic quark model, where X(6900) was suggested to be the candidate of the first radially excited tetraquarks with $J^{PC}$=0$^{++}$ or 2$^{++}$, or the 1$^{-+}$ or 2$^{-+}$ $P$-wave states. Besides, other interpretations, such as coupled-channel effects of double-charmonium channels\cite{Tc53,Tc54,Tc541,Tc542,Tc543}, $\mathrm{c}\bar{\mathrm{c}}$ hybrid\cite{Tc55}, and a Higgs-like boson\cite{Tc56} were also proposed. In summery, although many theoretical researches about the fully charmed tetraquark states have been reported, the interpretations for X(6900), X(6600), X(7200), etc. are still controversial.

 In our previous work, we used the relativistic quark model to study the mass spectra, r.m.s. radii and the radial density distributions of the singly and doubly charmed baryons. At present, we extend our previous method to analyze the fully charmed tetraquark states. We hope this study can help to shed more light on the nature of these new exotic states. The paper is organized as follows. After the introduction, we briefly describe the phenomenological method adopted in this work in Sec.II. In Sec.III we present our numerical results and discussions about the full-charmed tetraquark states. And Sec IV is reserved for our conclusions.

\begin{Large}
\textbf{2 Phenomenological method adopted in this work}
\end{Large}

The relativistic quark model has been successfully extended to study the mass spectrum of the $S$-wave tetraquark state\cite{Tc30}. In the following, we give a brief introduction to the Hamiltonian of relativized quark model.
The Hamiltonian for a four-body system is composed of the relativistic kinetic energy term, the confining potentials and one-gluon exchange potentials\cite{quark1,quark2},
\begin{eqnarray}
H=H_{0}+\sum_{i<j}V_{ij}^{\mathrm{conf}}+\sum_{i<j}V_{ij}^{\mathrm{oge}}
\end{eqnarray}
where the relativistic kinetic energy term is,
\begin{eqnarray}
H_{0}=\sum_{i=1}^{4}(p_{i}^{2}+m_{i}^{2})^{1/2}
\end{eqnarray}
The confining potential $V_{ij}^{\mathrm{conf}}$ is written as,
\begin{eqnarray}
V_{ij}^{\mathrm{conf}}=-\frac{3}{4}\textbf{\emph{F}}_{i}\cdot\textbf{\emph{F}}_{j}\Big[b r_{ij}\big[\frac{e^{-\sigma_{ij}^{2}r_{ij}^{2}}}{\sqrt{\pi}\sigma_{ij} r_{ij}}+\big(1+\frac{1}{2\sigma_{ij}^{2}r_{ij}^{2}}\big)\frac{2}{\sqrt{\pi}} \int^{\sigma_{ij} r_{ij}}_{0}e^{-x^{2}}dx\big]+c\Big]
\end{eqnarray}
with
\begin{eqnarray}
\sigma_{ij}=\sqrt{s^{2}\Big[\frac{2m_{i}m_{j}}{m_{i}+m_{j}}\Big]^{2}+\sigma_{0}^{2}\Big[\frac{1}{2}\big(\frac{4m_{i}m_{j}}{(m_{i}+m_{j})^{2}}\big)^{4}+\frac{1}{2}\Big]}
\end{eqnarray}
In Eq.(3), $\textbf{\emph{F}}_{i}\cdot\textbf{\emph{F}}_{j}$ stands for the color matrix and $\textbf{\emph{F}}$ reads
\begin{equation}
F_{n}=\left\{
      \begin{array}{l}
       \frac{\lambda_{n}}{2} \quad \mathrm{for} \, \mathrm{quarks}, \\
        -\frac{\lambda_{n}^{*}}{2} \quad    \mathrm{for} \, \mathrm{antiquarks} \\
      \end{array}
      \right.
\end{equation}
with $n=1,2\cdots8$.

The one-gluon
exchange potential $V_{ij}^{\mathrm{oge}}$ is composed of a Coulomb term $V_{ij}^{\mathrm{Coul}}$, hyperfine interaction $V^{\mathrm{hyp}}_{ij}$ and spin-orbital interaction $V^{\mathrm{so}}_{ij}$,
\begin{eqnarray}
V^{\mathrm{oge}}_{ij}=V_{ij}^{\mathrm{Coul}}+V^{\mathrm{hyp}}_{ij}+V^{\mathrm{so}}_{ij}
\end{eqnarray}
All of the one-gluon exchange potentials can be expressed in terms of a smeared one-gluon exchange propagator $\widetilde{G}$ which can be written as
\begin{eqnarray}
\widetilde{G}=\textbf{\emph{F}}_{i}\cdot\textbf{\emph{F}}_{j}\mathop{\sum}\limits_{k=1}^{3}\frac{2\alpha_{k}}{3\sqrt{\pi}r_{ij}}\int^{\tau_{k}r_{ij}}_{0}e^{-x^{2}}dx
\end{eqnarray}
where  $\tau_{k}=\frac{1}{\sqrt{\frac{1}{\sigma_{ij}^{2}}+\frac{1}{\gamma_{k}^{2}}}}$.

The Coulomb term $V_{ij}^{\mathrm{Coul}}$ is achieved by introducing a momentum-dependent factors $\Big(1+\frac{p^{2}_{ij}}{E_{i}E_{j}}\Big)^{\frac{1}{2}}$, and
\begin{eqnarray}
V_{ij}^{\mathrm{Coul}}=\Big(1+\frac{p^{2}_{ij}}{E_{i}E_{j}}\Big)^{\frac{1}{2}}\widetilde{G}(r_{ij})\Big(1+\frac{p^{2}_{ij}}{E_{i}E_{j}}\Big)^{\frac{1}{2}}
\end{eqnarray}
The hyperfine interaction $V^{\mathrm{hyp}}$ includes tensor interaction and the contact interaction,
\begin{eqnarray}
V^{\mathrm{hyp}}_{ij}=V^{\mathrm{tens}}_{ij}+V_{ij}^{\mathrm{cont}}
\end{eqnarray}
with
\begin{eqnarray}
V^{\mathrm{tens}}_{ij}=-\frac{1}{3m_{i}m_{j}}\Big(\frac{3\textbf{S}_{i}\cdot \textbf{r}_{ij}\textbf{S}_{j}\cdot \textbf{r}_{ij}}{r_{ij}^{2}}-\textbf{S}_{i}\cdot\textbf{S}_{j}\Big)\times\Big(\frac{\partial^{2}}{\partial r_{ij}^{2}}-\frac{1}{r_{ij}}\frac{\partial}{\partial r_{ij}}\Big)\widetilde{G}_{ij}^{\mathrm{t}},
\end{eqnarray}
\begin{eqnarray}
V^{\mathrm{cont}}_{ij}=\frac{2\textbf{S}_{i}\cdot\textbf{S}_{j}}{3m_{i}m_{j}}\bigtriangledown^{2}\widetilde{G}_{ij}^{\mathrm{c}}
\end{eqnarray}
For the spin-orbit interaction, it can also be divided into two parts which can be written as,
\begin{eqnarray}
V^{\mathrm{so}}_{ij}=V^{\mathrm{so(v)}}_{ij}+V_{ij}^{\mathrm{so(s)}},
\end{eqnarray}
with
\begin{eqnarray}
V^{\mathrm{so(v)}}_{ij}=\frac{\textbf{S}_{i}\cdot \textbf{L}_{ij}}{2m_{i}^{2}r_{ij}}\frac{\partial \widetilde{G}^{\mathrm{so(v)}}_{ii}}{\partial r_{ij}}+\frac{\textbf{S}_{j}\cdot \textbf{L}_{ij}}{2m_{j}^{2}r_{ij}}\frac{\partial \widetilde{G}^{\mathrm{so(v)}}_{jj}}{\partial r_{ij}}+\frac{(\textbf{S}_{i}+\textbf{S}_{j})\cdot \textbf{L}_{ij}}{m_{i}m_{j}r_{ij}}\frac{\partial \widetilde{G}^{\mathrm{so(v)}}_{ij}}{\partial r_{ij}}
\end{eqnarray}
and
\begin{eqnarray}
V^{\mathrm{so(s)}}_{ij}=-\frac{\textbf{S}_{i}\cdot \textbf{L}_{ij}}{2m_{i}^{2}r_{ij}}\frac{\partial \widetilde{V}^{\mathrm{so(s)}}_{ii}}{\partial r_{ij}}-\frac{\textbf{S}_{j}\cdot \textbf{L}_{ij}}{2m_{j}^{2}r_{ij}}\frac{\partial \widetilde{V}^{\mathrm{so(s)}}_{jj}}{\partial r_{ij}}
\end{eqnarray}
In Eqs.(10),(11),(13) and (14), $\widetilde{G}^{\mathrm{t}}_{ij}$, $\widetilde{G}^{\mathrm{c}}_{ij}$, $\widetilde{G}^{\mathrm{so(v)}}_{ij}$ and $\widetilde{V}^{\mathrm{\mathrm{so(s)}}}_{ii}$ are achieved from the $\widetilde{G}(r_{ij})$ and confining potential $V^{\mathrm{conf}}_{ij}(r_{ij})$ by introducing momentum-dependent factors,
\begin{eqnarray}
G^{\mathrm{t}}_{ij}=\Big(\frac{m_{i}m_{j}}{E_{i}E_{j}}\Big)^{\frac{1}{2}+\epsilon_{\mathrm{t}}}\widetilde{G}(r_{ij})\Big(\frac{m_{i}m_{j}}{E_{i}E_{j}}\Big)^{\frac{1}{2}+\epsilon_{\mathrm{t}}}
\end{eqnarray}
\begin{eqnarray}
G^{\mathrm{c}}_{ij}=\Big(\frac{m_{i}m_{j}}{E_{i}E_{j}}\Big)^{\frac{1}{2}+\epsilon_{\mathrm{c}}}\widetilde{G}(r_{ij})\Big(\frac{m_{i}m_{j}}{E_{i}E_{j}}\Big)^{\frac{1}{2}+\epsilon_{\mathrm{c}}}
\end{eqnarray}
\begin{eqnarray}
G^{\mathrm{so(v)}}_{ij}=\Big(\frac{m_{i}m_{j}}{E_{i}E_{j}}\Big)^{\frac{1}{2}+\epsilon_{\mathrm{so(v)}}}\widetilde{G}(r_{ij})\Big(\frac{m_{i}m_{j}}{E_{i}E_{j}}\Big)^{\frac{1}{2}+\epsilon_{\mathrm{so(v)}}}
\end{eqnarray}
\begin{eqnarray}
\widetilde{V}^{\mathrm{so(s)}}_{ii}=\Big(\frac{m_{i}^{2}}{E_{i}^{2}}\Big)^{\frac{1}{2}+\epsilon_{\mathrm{so(s)}}}V_{ij}^{\mathrm{conf}}(r_{ij})\Big(\frac{m_{i}^{2}}{E_{i}^{2}}\Big)^{\frac{1}{2}+\epsilon_{\mathrm{so(s)}}}
\end{eqnarray}
with $E_{i}=\sqrt{m_{i}^{2}+p_{ij}^{2}}$, where $p_{ij}$ is the magnitude of the momentum of either of the quarks in the $ij$ center-of-mass frame.

The internal motions of the quarks in a four-body system can be expressed by three sets of Jacobi coordinates as shown in Fig.1. As for the Jacobi coordinates in Fig.1(a), they can be defined as,
\begin{eqnarray}
& \boldsymbol{r}_{12}=\textbf{\emph{r}}_{2}-\textbf{\emph{r}}_{1}& \\
& \boldsymbol{r}_{34}=\textbf{\emph{r}}_{4}-\textbf{\emph{r}}_{3} & \\
& \boldsymbol{r}=\frac{\textbf{\emph{r}}_{4}+\textbf{\emph{r}}_{3}}{2}-\frac{\textbf{\emph{r}}_{1}+\textbf{\emph{r}}_{2}}{2}& \\
& \boldsymbol{R}=\frac{\textbf{\emph{r}}_{1}+\textbf{\emph{r}}_{2}+\textbf{\emph{r}}_{3}+\textbf{\emph{r}}_{4}}{4}
\end{eqnarray}
\begin{figure}[h]
\centering
\includegraphics[height=4.5cm,width=14cm]{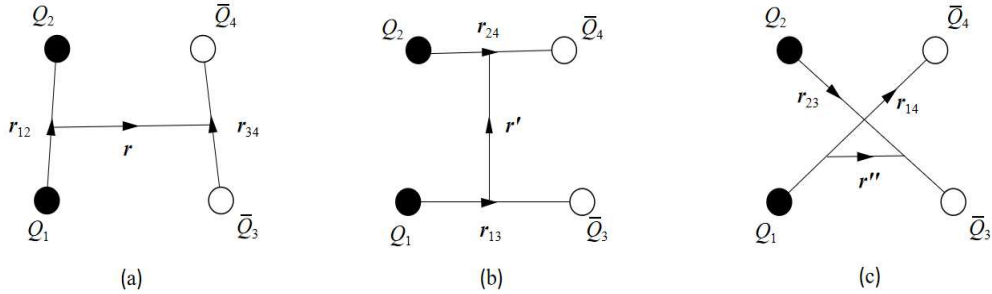}
\caption{ Jacobi coordinates for the four-body system.}
\end{figure}
According to the Jacobi transformation, the other two sets of coordinates can also be expressed in terms of $\textbf{\emph{r}}_{12}$, $\textbf{\emph{r}}_{34}$, and $\textbf{\emph{r}}$.
The wave function for a fully charmed tetraquark state is composed of color, flavor, spin, and spatial parts. The calculations in this work are based on the Jacobi coordinates in Fig.1(a).
Under this picture, the colorless wave function can be expressed as $|(Q_{1}Q_{2})_{\bar{3}}(\bar{Q}_{3}\bar{Q}_{4})_{3}\rangle$ and $|(Q_{1}Q_{2})_{6}(\bar{Q}_{3}\bar{Q}_{4})_{\bar{6}}\rangle$ which are antisymmetric and symmetric under the exchange of $Q_{1}Q_{2}$ or $\bar{Q}_{3}\bar{Q}_{4}$, respectively. In the flavor space, the fully charmed tetraquark state $[Q_{1}Q_{2}][\bar{Q}_{3}\bar{Q}_{4}]$ is always symmetric, where the square bracket denotes the flavor symmetry. For a double quark(untiquark) system in the fully charmed tetraquark state, its spin wave function is antisymmetric singlet or symmetric triplet and they can be expressed as $[Q_{1}Q_{2}]^{0}$, $[\bar{Q}_{3}\bar{Q}_{4}]^{0}$ and $[Q_{1}Q_{2}]^{1}$, $[\bar{Q}_{3}\bar{Q}_{4}]^{1}$, respectively.

In this work, the Gaussian basis function is employed to construct the spatial wave function of the tetrquark state, and the Gaussian basis reads
\begin{eqnarray}
\phi_{nlm_{l}}(\boldsymbol{r})=N_{nl}r^{l}e^{-\nu_{n}r^{2}}Y_{lm_{l}}(\hat{\boldsymbol{r}})
\end{eqnarray}
with
\begin{eqnarray}
N_{nl}=\sqrt{\frac{2^{l+2}(2\nu_{n})^{l+3/2}}{\sqrt{\pi}(2l+1)!!}}
\end{eqnarray}
\begin{eqnarray}
\nu_{n}=\frac{1}{r_{n}^{2}}, \quad r_{n}=r_{a1}\Big[\frac{r_{n_{\mathrm{max}}}}{r_{a 1}}\Big]^{\frac{n-1}{n_{\mathrm{max}}-1}}
\end{eqnarray}
In Eq.(25), $n_{max}$ is the maximum number of the Gaussian basis functions. The total wave function of the spin and spatial parts with the angular momentum ($J$,$M$) can be written as,
\begin{eqnarray}
\psi_{JM}=\sum_{\kappa}C_{\kappa}[[(\phi_{n_{12}l_{12}}(\mathbf{r}_{12})\otimes\chi_{s_{12}})_{j_{a}}(\phi_{n_{34}l_{34}}(\mathbf{r}_{34})\otimes\chi_{s_{34}})_{j_{b}}]_{j}\otimes\phi_{nl}(\mathbf{r})]_{JM}
\end{eqnarray}
where $\kappa$ is the quantum numbers $\{$$n_{12}$,$l_{12}$,$s_{12}$,$j_{a}$,$n_{34}$,$l_{34}$,$s_{34}$,$j_{b}$,$n$,$l$,$j$$\}$ of the basis. Theoretically, a tetraquark state with total angular momentum $J$ is the superposition of all the bases. The basis can be classified by the total orbital angular momentum $L$=$l_{12}$+$l_{34}$+$l$. The tetraquark states with different orbital angular momentum $L$ but with same $J^{PC}$ will mix with each other, for example, the $0^{++}$ state of $L$=0 will couple with $L$=2, and $L$=4 states through spin-orbital and tensor potentials. This interaction will influence the mass spectrum slightly, thus we neglect this mixing mechanism in this work.

According to the Pauli exclusion principle, the total wave function of a tetraquark should be antisymmetric, and all possible configurations for
fully charmed tetraquark are presented in the third column of Table IV. For a $S$-wave tetraquark state, its possible spin and parity quantum numbers are $J^{PC}$=$0^{++}$, $1^{+-}$, and $2^{++}$. For the $P$-wave states, there are two orbital excited modes, the $\rho$-mode with the orbital excitation in the diquark or antidiquark, i.e., ($l_{12}$,$l_{34}$,$l$)=(1,0,0) or (0,1,0), and the $\lambda$-mode with the orbital excitation between the two clusters, i.e., ($l_{12}$,$l_{34}$,$l$)=(0,0,1). Besides of the conventional quantum numbers, i.e., $J^{PC}=0^{-+}$, $1^{--}$, $2^{-+}$, $3^{--}$, the $P$-wave can also access exotic quantum numbers, i.e., $J^{PC}=0^{--}$, $1^{-+}$, $2^{--}$. For simplicity, each configuration in Table IV can be expressed as $|c_{\rho/\lambda}^{C};^{2S+1}L_{J}\rangle$, where $c$=3 or 6 stand for the color configuration, $C$ and $S$ are the C-parity and the total spin angular momentum of the configuration.

For a four-body system, the calculations of the Hamiltonian matrix elements become laborious even with Gaussian basis functions.
This process can be simplified by introducing the ISG basis functions. These new sets of basis functions can be written as\cite{ISG},
\begin{eqnarray}
\phi_{nlm_{l}}(\boldsymbol{r})=N_{nl}\lim_{\varepsilon\rightarrow 0}\frac{1}{(\nu_{n}\varepsilon)^{l}}\sum_{k=1}^{k_{\mathrm{max}}}C_{lm_{l},k}e^{-\nu_{n}(\textbf{r}-\varepsilon \textbf{D}_{lm_{l},k})^{2}}
\end{eqnarray}
where $\varepsilon$ is the shifted distance of the Gaussian basis. Taking the limit $\varepsilon\rightarrow 0$ is to be carried out after the matrix elements have been calculated analytically. For more details about the calculations of the Hamiltonian matrix elements, one can consult the Refs.\cite{ISG,GLY1}.

After all of the matrix elements are evaluated, the mass spectra can be obtained by solving the generalized eigenvalue problem,
\begin{flalign}
\sum_{j=1}^{n_{\mathrm{max}}^{3}}\Big(H_{ij}-E\widetilde{N}_{ij}\Big)C_{j}=0, \quad (i=1-n_{\mathrm{max}}^{3})
\end{flalign}
Here, $H_{ij}$ denotes the matrix element in the total color-flavor-spin-spatial base, $E$ is the eigenvalue, $C_{j}$ stands for the corresponding eigenvector, and $\widetilde{N}_{ij}$ is the overlap matrix elements of the Gaussian functions, which arises from the nonorthogonality of the bases and can be expressed as,
\begin{flalign}
\notag
\widetilde{N}_{ij}\equiv \langle\phi_{n_{12}l_{12}m_{l_{12}}}|
\phi_{n_{12}^{\prime}l_{12}^{\prime}m_{l_{12}^{\prime}}}\rangle\times
\langle\phi_{n_{34}l_{34}m_{l_{34}}}
|\phi_{n_{34}^{\prime}l_{34}^{\prime}m_{l_{34}^{\prime}}}\rangle\times\langle\phi_{nlm_{l}}|
\phi_{n^{\prime}l^{\prime}m_{l^{\prime}}}\rangle & \\
=\Big(\frac{2\sqrt{\nu_{n_{12}}\nu_{n_{12}^{\prime}}}}{\nu_{n_{12}}+\nu_{n_{12}^{\prime}}}\Big)^{l_{12}+3/2}\times
\Big(\frac{2\sqrt{\nu_{n_{34}}\nu_{n_{34}^{\prime}}}}{\nu_{n_{34}}+\nu_{n_{34}^{\prime}}}\Big)^{l_{34}+3/2}\times\Big(\frac{2\sqrt{\nu_{n}\nu_{n^{\prime}}}}{\nu_{n}+\nu_{n^{\prime}}}\Big)^{l+3/2} &
\end{flalign}

\begin{Large}
\textbf{3 Numerical results and discussions}
\end{Large}

\begin{large}
\textbf{3.1 S- and P-wave tetraquark states}
\end{large}

\begin{table*}[h]
\begin{ruledtabular}\caption{Relevant parameters of the relativized quark model}
\begin{tabular}{c c c c c }
$m_{c}$(GeV)&$\alpha_{1}$ & $\alpha_{2}$& $\alpha_{3}$ &$\gamma_{1}$(GeV) \\
$1.628$&  $0.25$ & $0.15$&$0.20$ &$\frac{1}{2}$  \\ \hline
$\gamma_{2}$(GeV)& $\gamma_{3}$(GeV) &$b$(GeV$^{2}$)& $c$(GeV)&$\sigma_{0}$(GeV) \\
$\sqrt{10}/2$& $\sqrt{1000}/2$ & $0.18$ &$-0.253$ &$1.8$\\ \hline
$s$&$\epsilon_{\mathrm{c}}$ & $\epsilon_{\mathrm{so(v)}}$& $\epsilon_{\mathrm{t}}$ &$\epsilon_{\mathrm{so(s)}}$ \\
 $1.55$ & $-0.168$ & $-0.035$ &$0.025$ &$0.055$ \\
\end{tabular}
\end{ruledtabular}
\end{table*}
The results of the relativized quark model depend on the input parameters such as the constituent quark mass and the parameters in the Hamiltonian. Up to now, there has been no solid experimental data for the spectrum of pure tetraquark states, it is impossible for us to fix these parameters by fitting present experimental data. In most cases, these parameters were determined by fitting them to the experimental masses of the mesons or baryons\cite{Tc26,Tc29,Tc31,para1,para2,para3}. In the present work, all input parameters are listed in Table II and taken from the original Refs.\cite{quark1,quark2} where they successfully reproduced the excremental data of mesons and baryons. In our previous work, predictions for heavy baryons with these parameters were indeed consistent well with the experimental data\cite{GLY1}. Recently, these parameters were already extended to study the tetraquark states in a uniform frame. As for the accuracy of the relativized quark model, it depends on the quenched approximation and relativistic corrections. Considering these two effects in Ref.\cite{quark1}, they claimed that the average accuracies are 25 MeV for light and heavy-light mesons and 10 MeV for heavy mesons, respectively. We expect that the uncertainties of predicted masses of the teatraquark states are limited in a reasonable range.

 In order to investigate the convergence and stability of the numerical results, we plot the masses of the configurations $|3^{-};^{3}S_{1}\rangle$ and $|3^{+}_{\lambda};^{3}P_{0}\rangle$ in Fig.2(a) and their r.m.s. radii $\sqrt {\langle {r_{12}^{2}}\rangle }$ in Fig.2(b). In these figures, the wave function is expanded with Gaussian basis $n_{\mathrm{max}}^{3}$=$2^{3}$, $3^{3}$, $4^{3}$, $5^{3}$, $6^{3}$, $7^{3}$, and $8^{3}$, respectively. We can see that the results decrease with the basis number and converge to a stable value when $n_{\mathrm{max}}^{3}>6^3$. In this work, the calculations are carried out with $n_{\mathrm{max}}^{3}=8^3$ Gaussian basis to ensure the stability of the final results.
\begin{figure}[H]
  \centering
   \subfigure[]{
   \begin{minipage}{5.5cm}
   \centering
   \includegraphics[width=6cm]{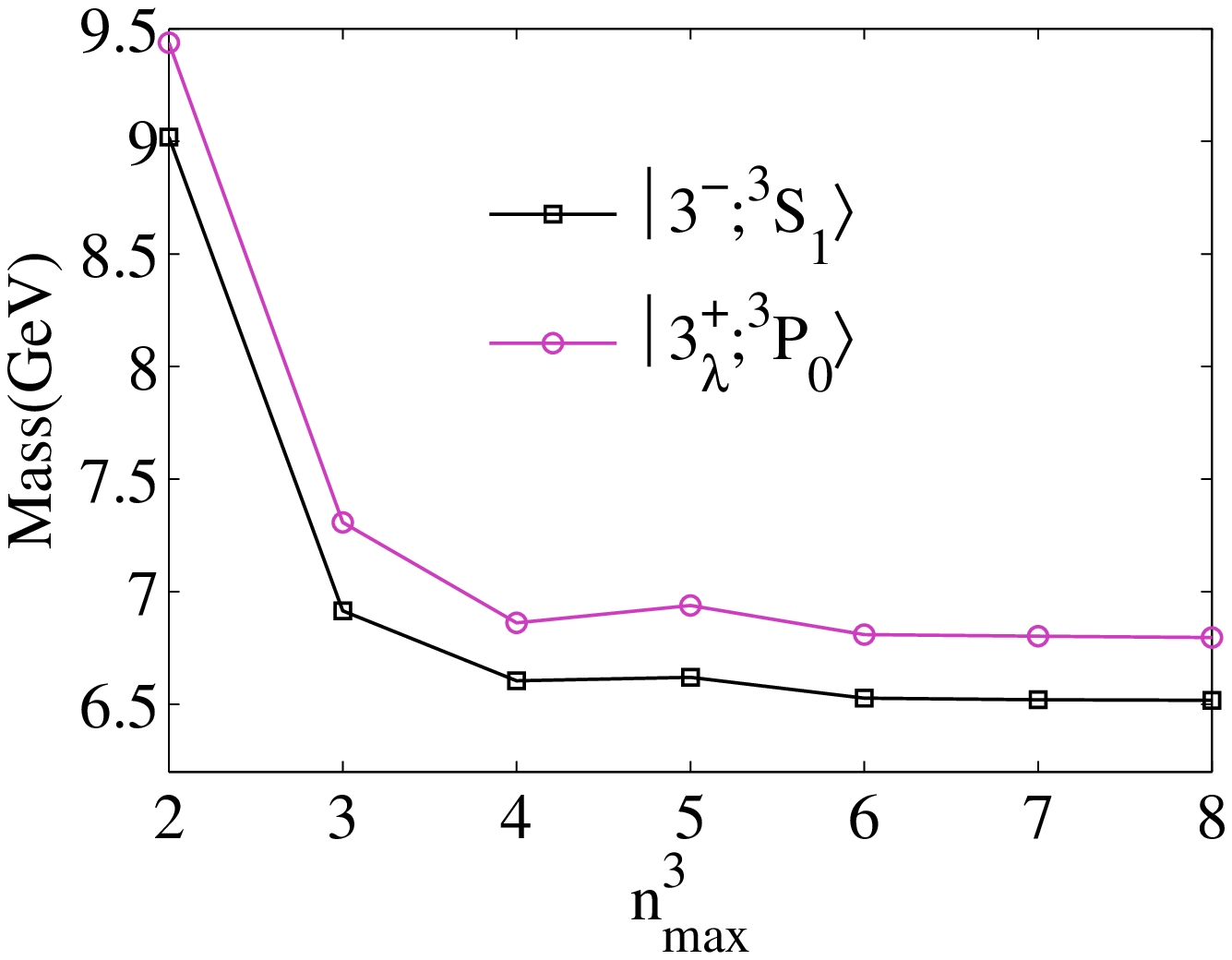}
  \end{minipage}
  }
 \subfigure[]{
   \begin{minipage}{5.5cm}
   \centering
   \includegraphics[width=6cm]{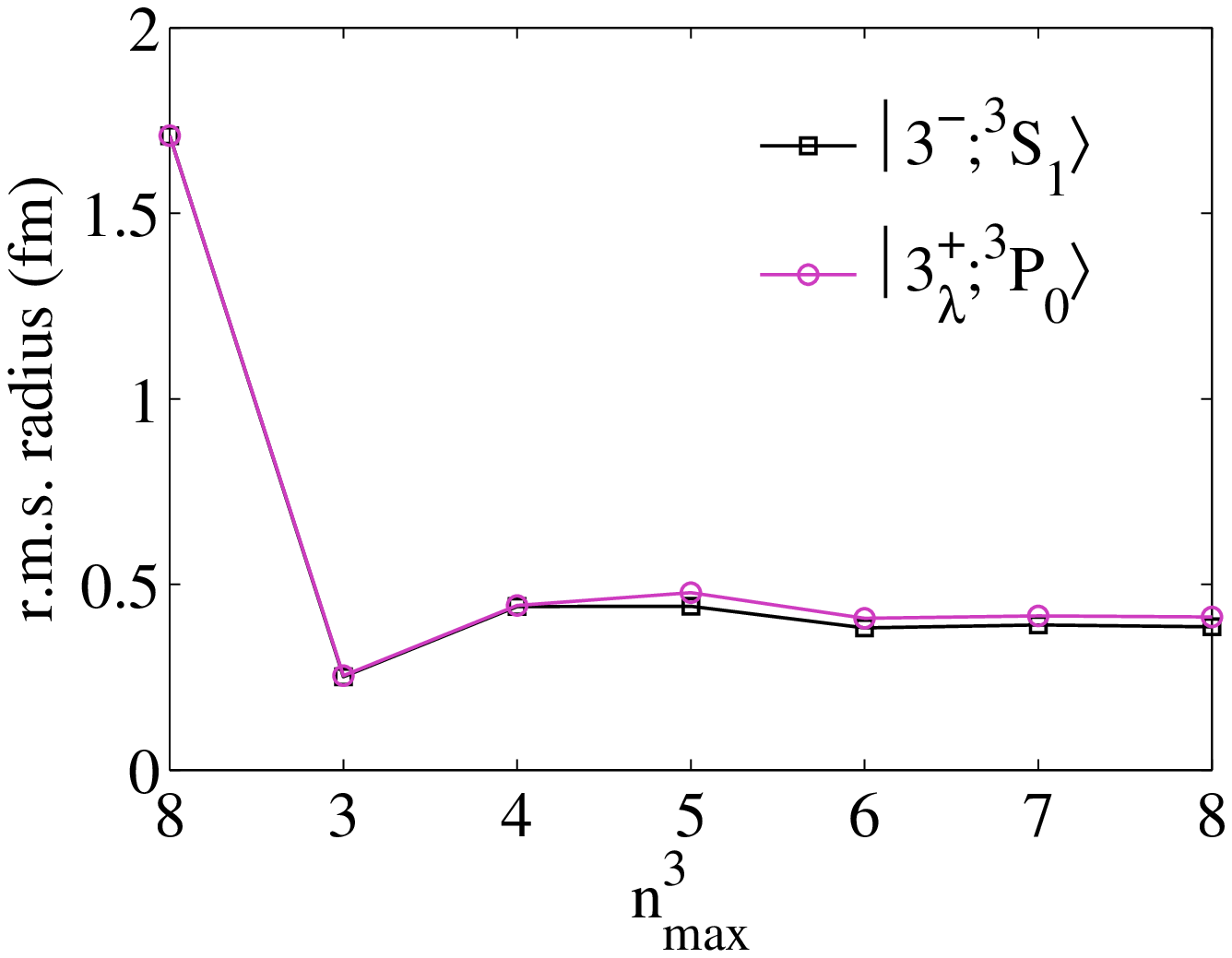}
  \end{minipage}
  }
  \caption{The dependence of the mass and r.m.s. radius on the number of
Gaussian basis.}
\end{figure}
In general, the physical state of a tetraquark are the mixtures of different configurations with the same quantum number of $J^{PC}$. Thus, the calculations are carried out in two stages. In the first stage, the masses of different configurations shown in the third column in Table IV are obtained by solving the Schr$\ddot{\mathrm{o}}$dinger equation with the variational method. The masses and the r.m.s. radii for the ground and the first radially excited states are also presented in Table IV. In the second stage, the mixing effect is considered and the masses of physical states are obtained by diagonalizing the mass matrix in the basis of eigenstates obtained in the first stage. The mass matrix, the eigenvalues, and the eigenvectors for the ground states and the first radial excitations are summarized in Tables V-VI, respectively.

In the OGE model, the interaction between the two quarks within a color-sextet diquark is repulsive, while that in the color-anti-triplet one is attractive. On the other hand, the interaction between the diquark and antidiquark of $|(Q_{1}Q_{2})_{6}(\bar{Q}_{3}\bar{Q}_{4})_{\bar{6}}\rangle$ configuration is attractive and is much stronger than that of the $|(Q_{1}Q_{2})_{\bar{3}}(\bar{Q}_{3}\bar{Q}_{4})_{3}\rangle$ one. From Table IV, one can see that $|6^{+};^{1}S_{0}\rangle$ configuration is located lower than the $|3^{+};^{1}S_{0}\rangle$. This can be explained by the stronger attractive potential between diquark and antidiquark in $|6^{+}\rangle$ configuration. This also applies to the $\rho$-mode excitations, i.e., 6$_{\rho}<$3$_{\rho}$. If the $\lambda$-mode excitations are also considered, the relationship is slightly complicated, where the relationship between different color configurations is 6$_{\rho}<$3$_{\lambda}<$3$_{\rho}<$6$_{\lambda}$, which can be seen in Table IV. This is the result of the competition of various interactions among quarks.

From Table IV, it is shown that the r.m.s. radii $\sqrt{\langle r_{12/34}^{2}\rangle}$ of $|3^{+}\rangle$ and $|3^{\pm}_{\rho}\rangle$ states are smaller than those of the $|6^{+}\rangle$ and $|6^{\pm}_{\rho}\rangle$, respectively. This is also due to the attractive interactions between the two quarks within a color-antitriplet diquark and repulsive ones in a color-sextet diquark. On the other hand, the stronger attraction between diquark and antidiquark in $|(Q_{1}Q_{2})_{6}(\bar{Q}_{3}\bar{Q}_{4})_{\bar{6}}\rangle$ configuration makes the situation of $\sqrt{\langle r^{2}\rangle}$ being opposite to $\sqrt{\langle r_{12/34}^{2}\rangle}$.
For example, $\sqrt{\langle r_{12/34}^{2}\rangle}$ and $\sqrt{\langle r^{2}\rangle}$ are 0.383 fm and 0.309 fm for the ground state of $|3^{+}\rangle$ configuration, while the results are 0.456 fm and 0.235 fm for $|6^{+}\rangle$ one.
To further understand the inner structures of different configurations, we also plot the radial density distributions which are obtained by the wave functions from quark model. The distribution functions are defined as,
\begin{eqnarray}
\notag
&\omega(r_{12/34})=\int|\Psi(\textbf{r}_{12},\textbf{r}_{34},\textbf{r})|^{2}d\textbf{r}d\textbf{r}_{34/12}d\Omega_{12/34} \\
&\omega(r)=\int|\Psi(\textbf{r}_{12},\textbf{r}_{34},\textbf{r})|^{2}d\textbf{r}_{12}d\textbf{r}_{34}d\Omega
\end{eqnarray}
where $\Omega_{12/34}$ and $\Omega$ are the solid angles spanned by vectors $\textbf{r}_{12/34}$ and $\textbf{r}$, respectively. Some of the results are shown in Figs. 3-8.
It can be seen that the radial density distributions are in the range of 1 fm, which indicates that the four quarks are confined into a compact state. Second, the peaks for the first radial excitations are located more outward compared with their ground states. Third, for the $\lambda$-mode excitation, e.g. $|3^{+}_{\lambda};^{3}P_{0}\rangle$, it is shown in Fig.6 that the $r^{2}\omega(r)$ peak is located more outward than that of $r^{2}_{12/34}\omega(r_{12/34})$, while the situation is opposite for $\rho$-mode excitation, e.g. $|3^{+}_{\rho};^{3}P_{0}\rangle$ in Fig.7.
\begin{figure}[H]
  \centering
   \subfigure[]{
   \begin{minipage}{5.5cm}
   \centering
   \includegraphics[width=6cm]{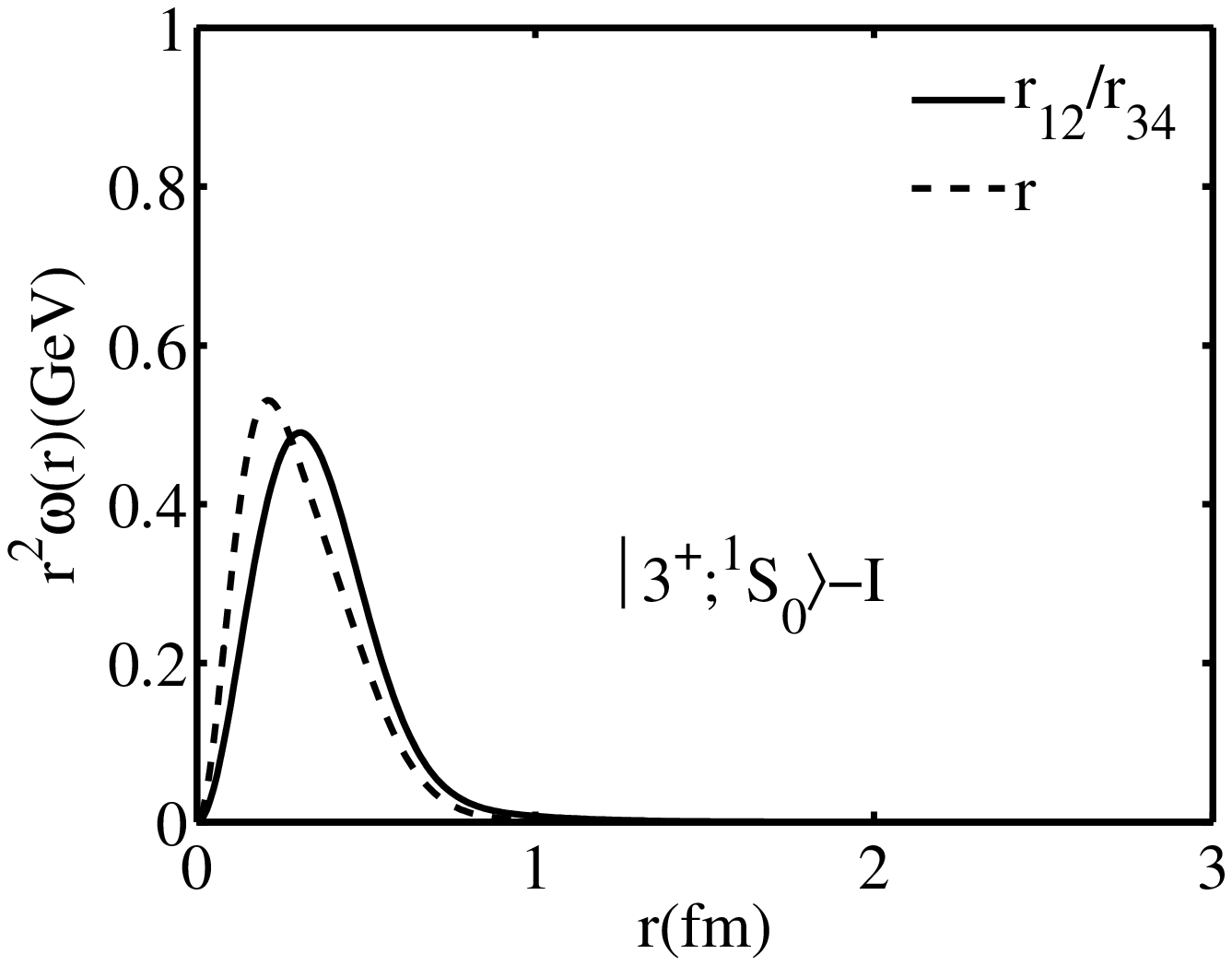}
  \end{minipage}
  }
 \subfigure[]{
   \begin{minipage}{5.5cm}
   \centering
   \includegraphics[width=6cm]{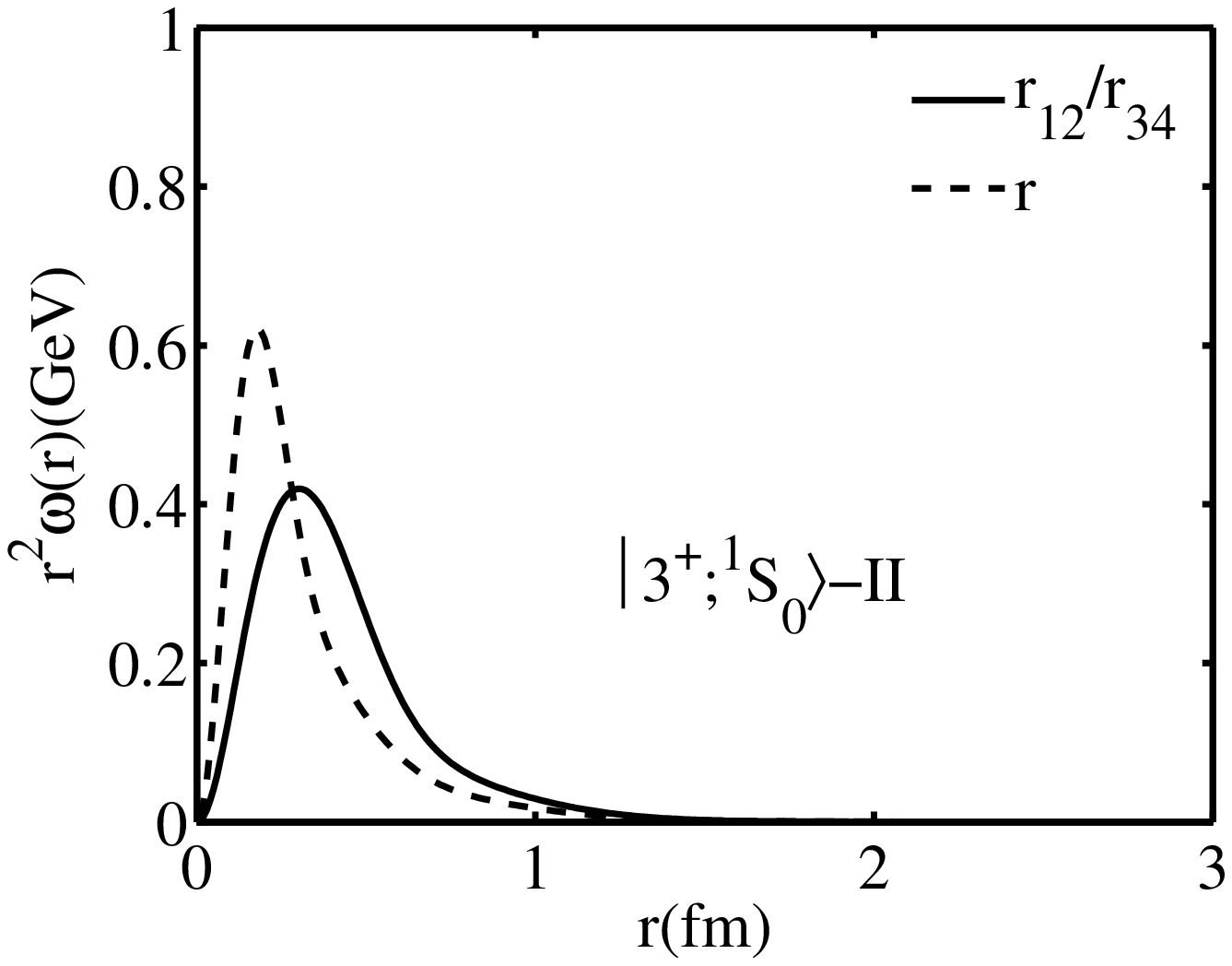}
  \end{minipage}
  }
  \caption{Radial density distributions of the ground state(a) and the first radially excited state(b) of $0^{++}$($|3^{+};^{3}S_{0}\rangle$)}
\end{figure}
\begin{figure}[H]
  \centering
   \subfigure[]{
   \begin{minipage}{5.5cm}
   \centering
   \includegraphics[width=6cm]{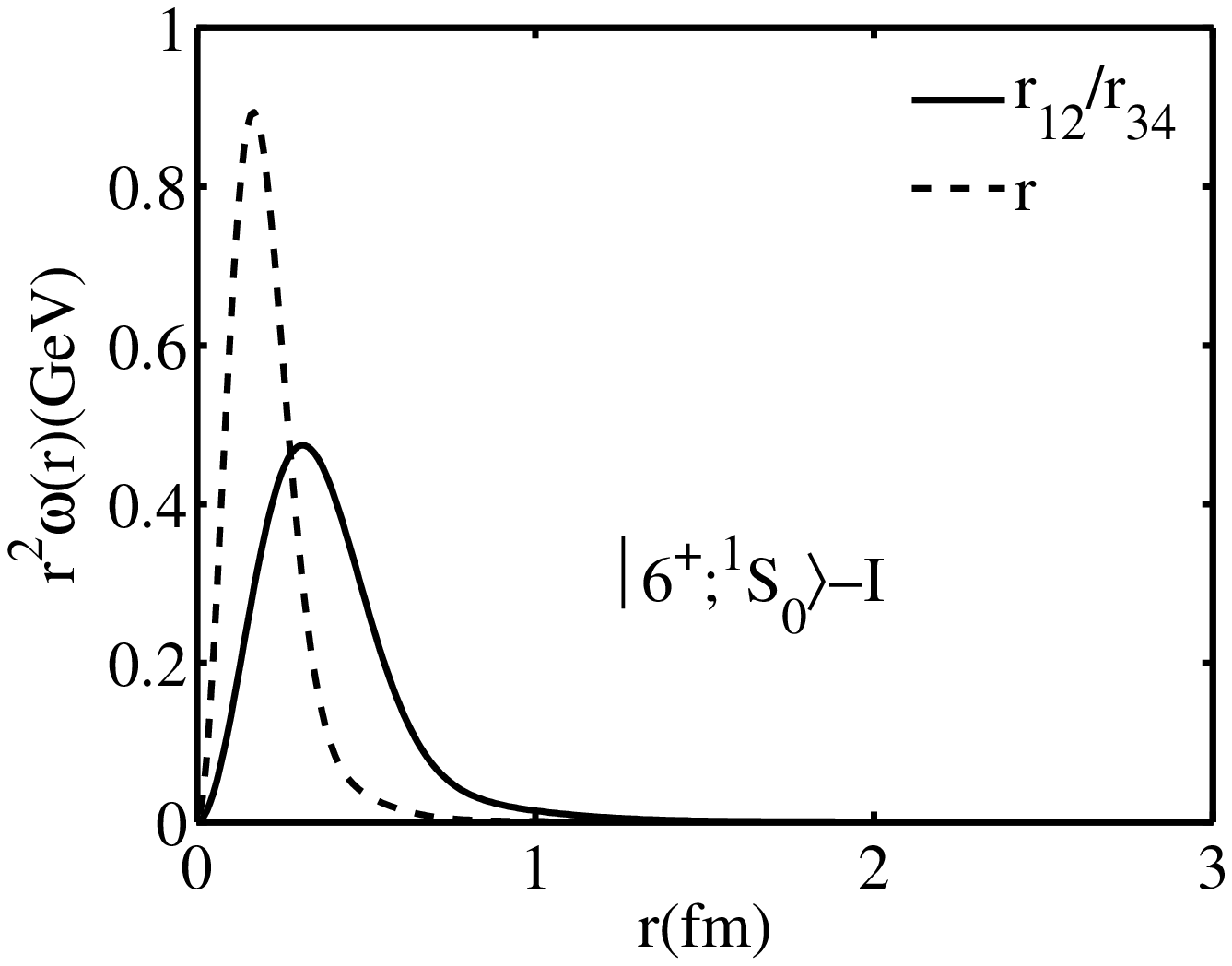}
  \end{minipage}
  }
 \subfigure[]{
   \begin{minipage}{5.5cm}
   \centering
   \includegraphics[width=6cm]{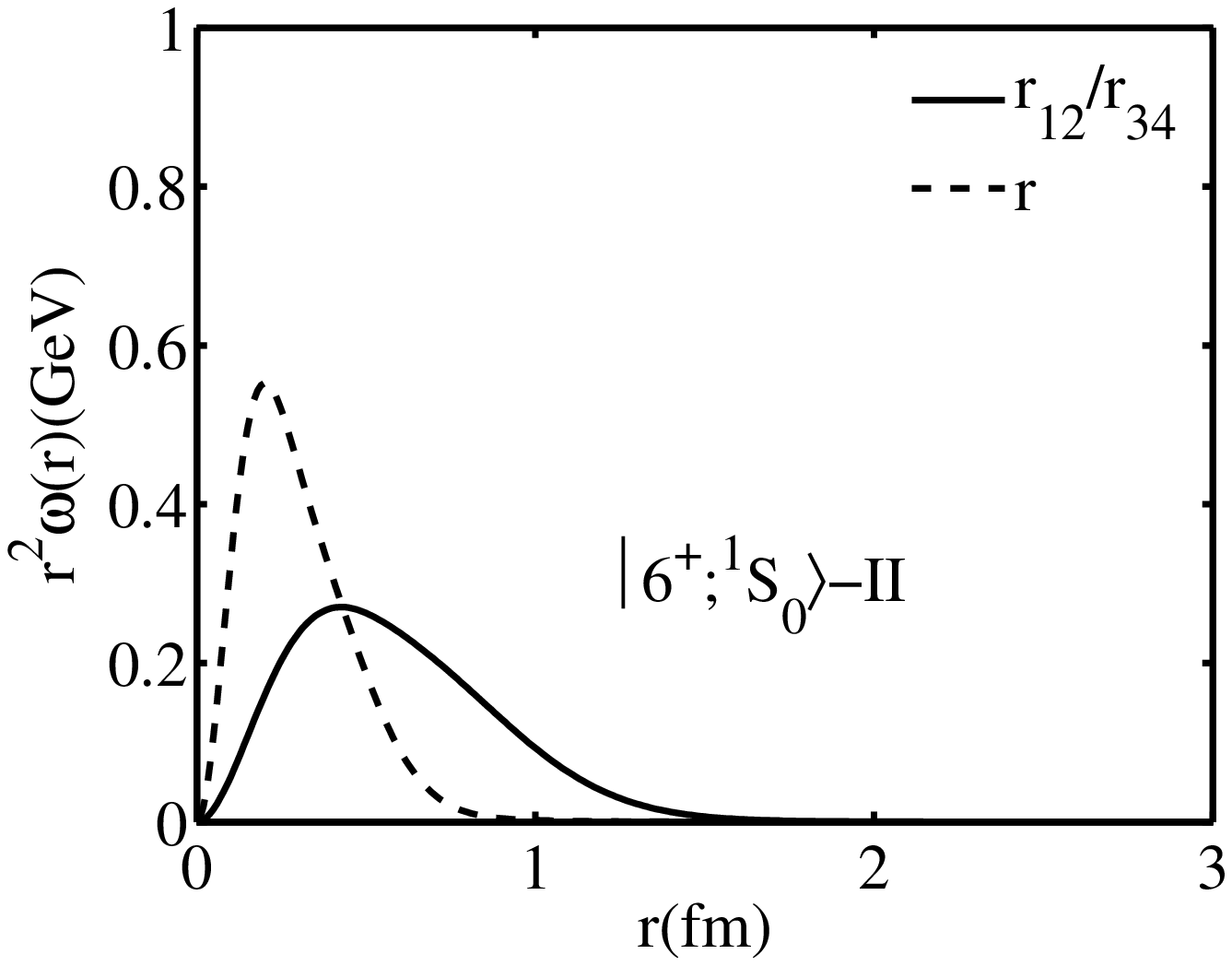}
  \end{minipage}
  }
  \caption{Radial density distributions of the ground state(a) and the first radially excited state(b) of $0^{++}$($|6^{+};^{3}S_{0}\rangle$)}
\end{figure}
\begin{figure}[H]
  \centering
   \subfigure[]{
   \begin{minipage}{5.5cm}
   \centering
   \includegraphics[width=6cm]{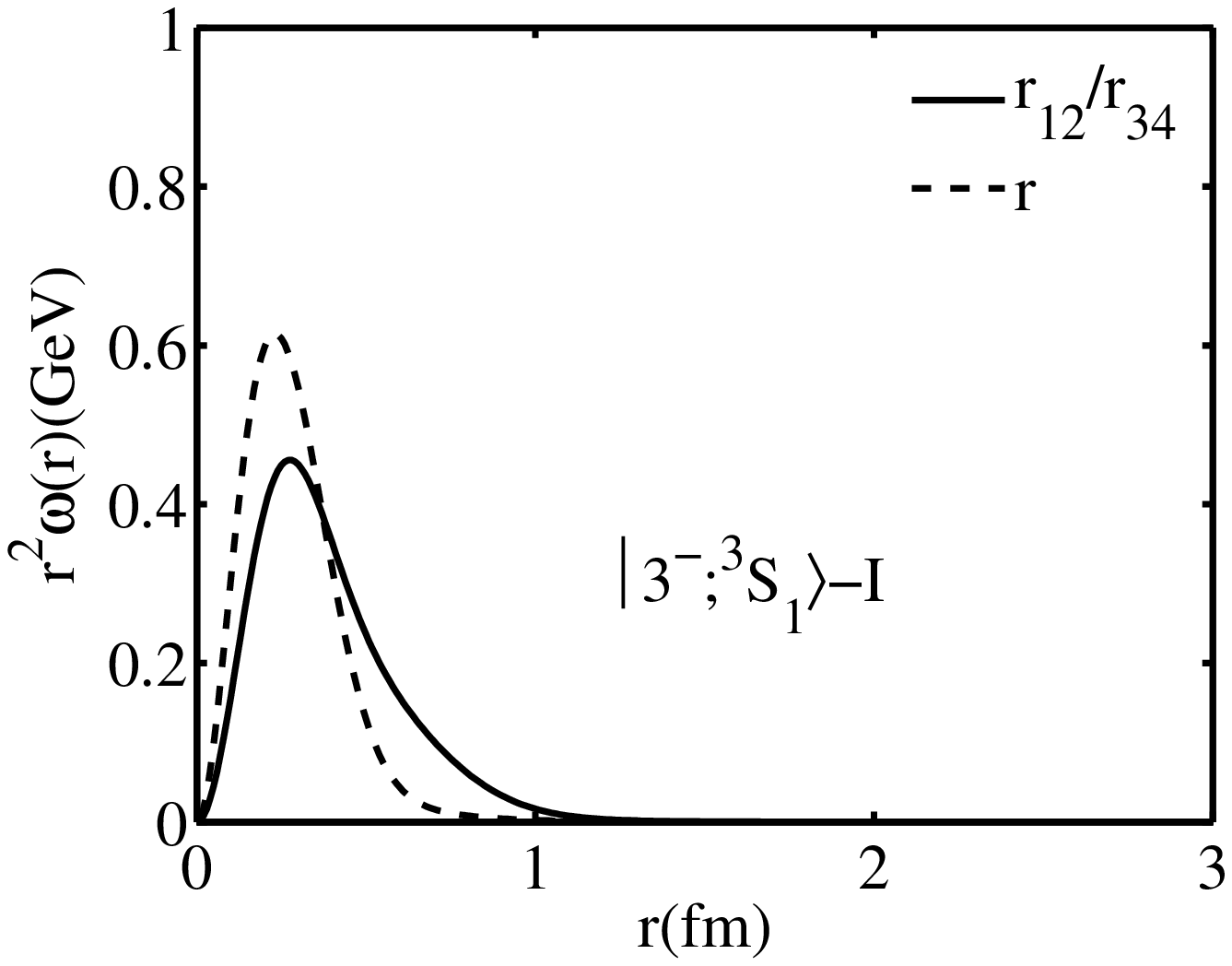}
  \end{minipage}
  }
 \subfigure[]{
   \begin{minipage}{5.5cm}
   \centering
   \includegraphics[width=6cm]{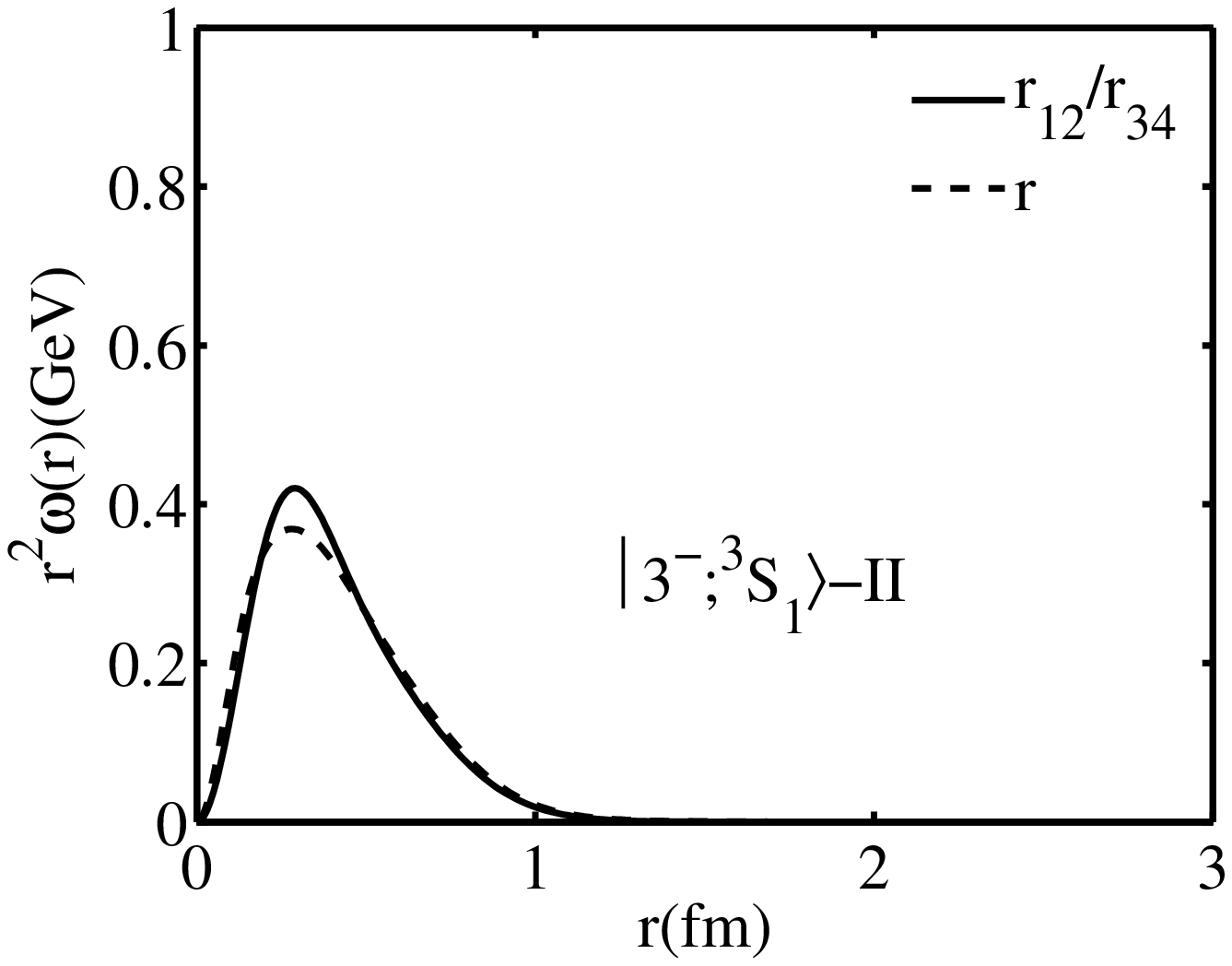}
  \end{minipage}
  }
  \caption{Radial density distributions of the ground state(a) and the first radially excited state(b) of $1^{+-}$($|3^{-};^{3}S_{1}\rangle$)}
\end{figure}
\begin{figure}[H]
  \centering
   \subfigure[]{
   \begin{minipage}{5.5cm}
   \centering
   \includegraphics[width=6cm]{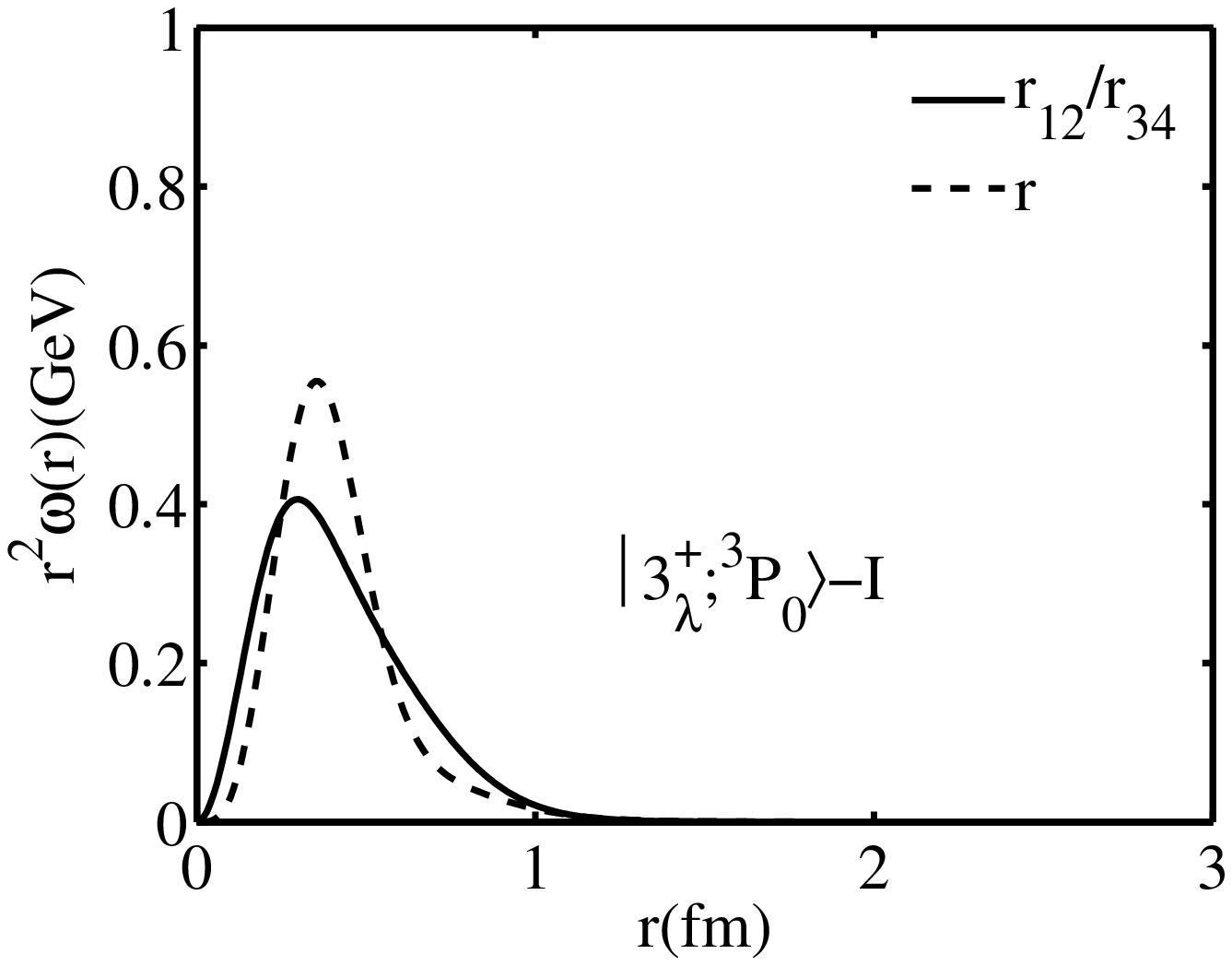}
  \end{minipage}
  }
 \subfigure[]{
   \begin{minipage}{5.5cm}
   \centering
   \includegraphics[width=6cm]{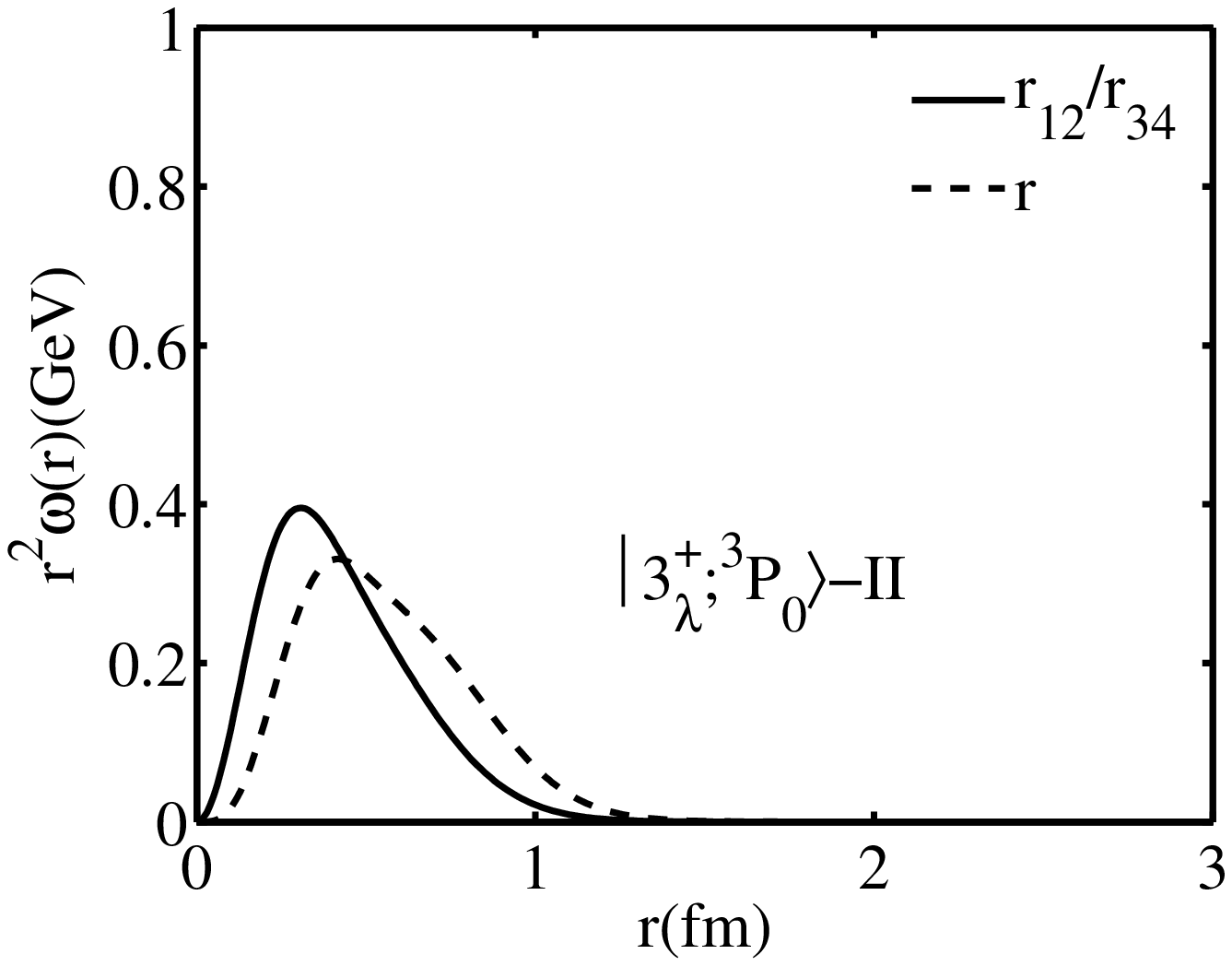}
  \end{minipage}
  }
  \caption{Radial density distributions of the ground state(a) and the first radially excited state(b) of $0^{-+}$($|3^{+}_{\lambda};^{3}P_{0}\rangle$)}
\end{figure}
\begin{figure}[H]
  \centering
   \subfigure[]{
   \begin{minipage}{5.5cm}
   \centering
   \includegraphics[width=6cm]{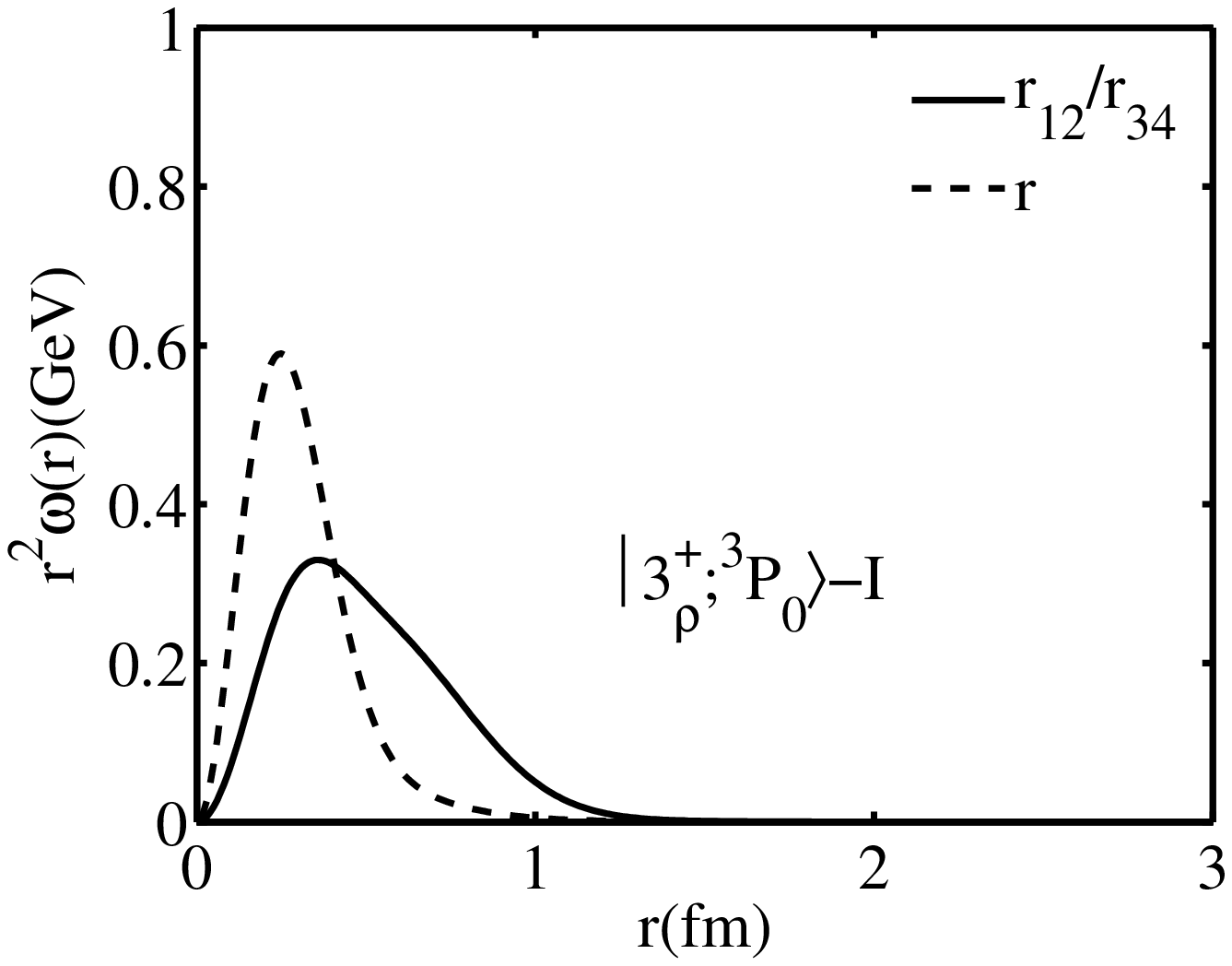}
  \end{minipage}
  }
 \subfigure[]{
   \begin{minipage}{5.5cm}
   \centering
   \includegraphics[width=6cm]{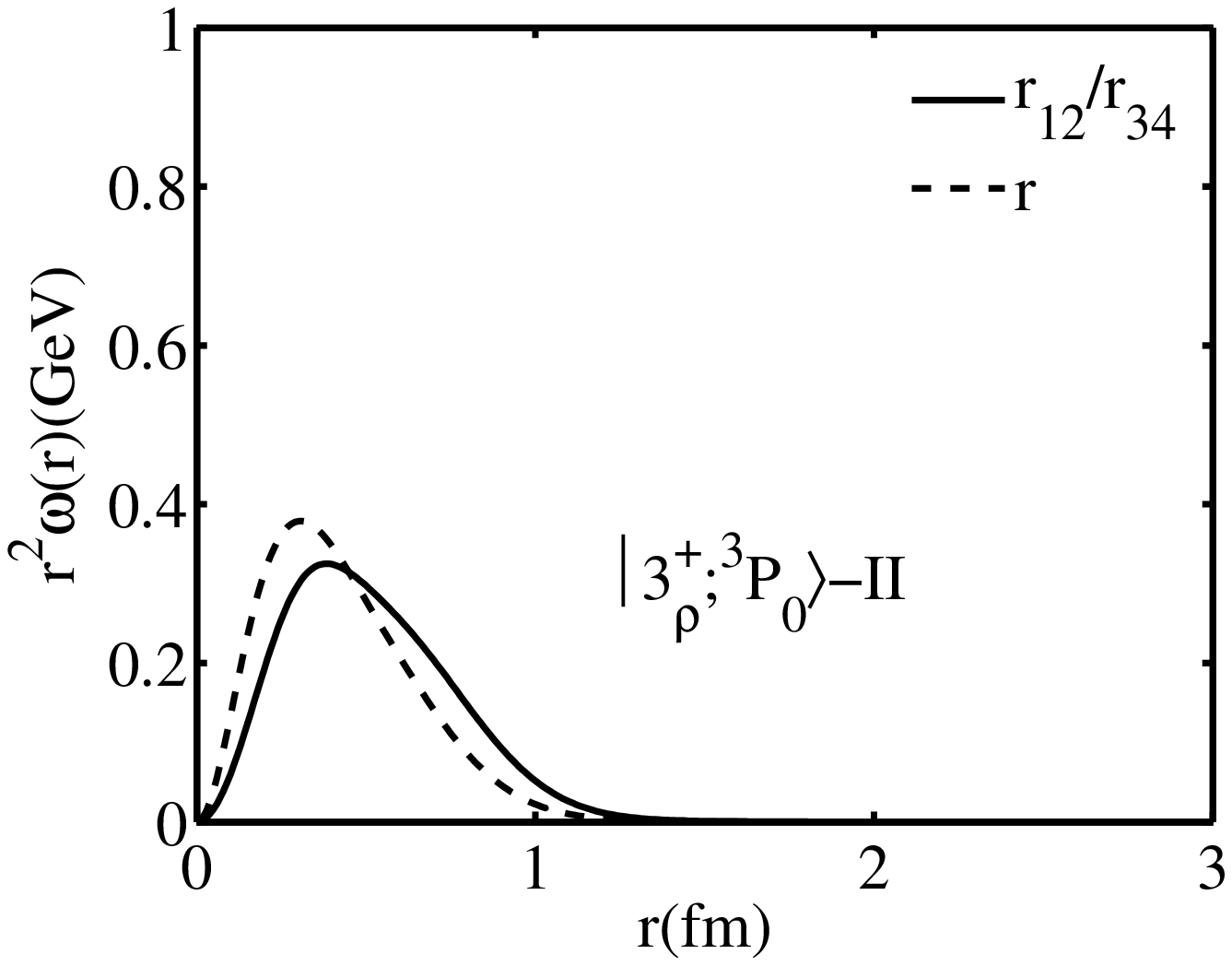}
  \end{minipage}
  }
  \caption{Radial density distributions of the ground state(a) and the first radially excited state(b) of $0^{-+}$($|3^{+}_{\rho};^{3}P_{0}\rangle$)}
\end{figure}
\begin{figure}[H]
  \centering
   \subfigure[]{
   \begin{minipage}{5.5cm}
   \centering
   \includegraphics[width=6cm]{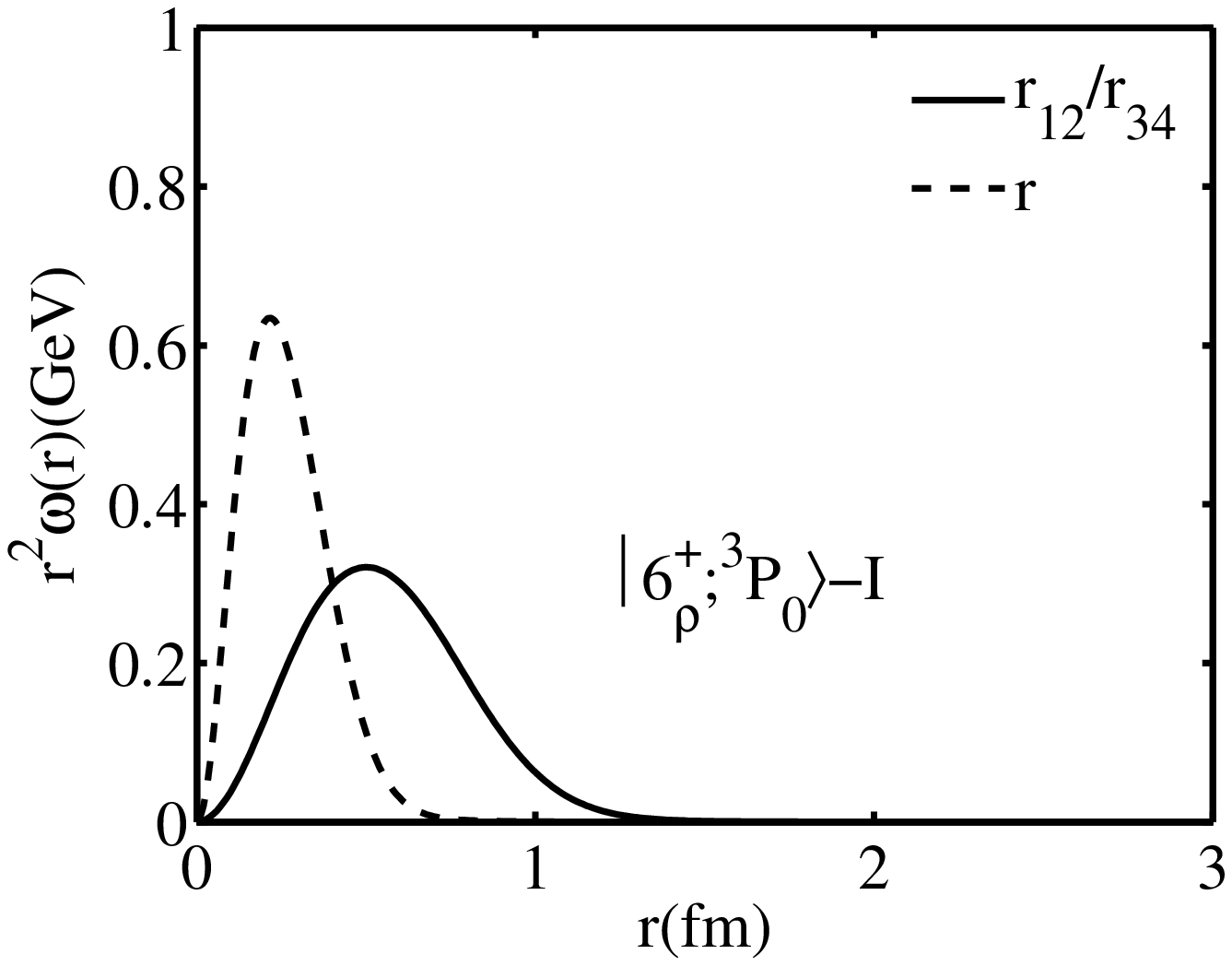}
  \end{minipage}
  }
 \subfigure[]{
   \begin{minipage}{5.5cm}
   \centering
   \includegraphics[width=6cm]{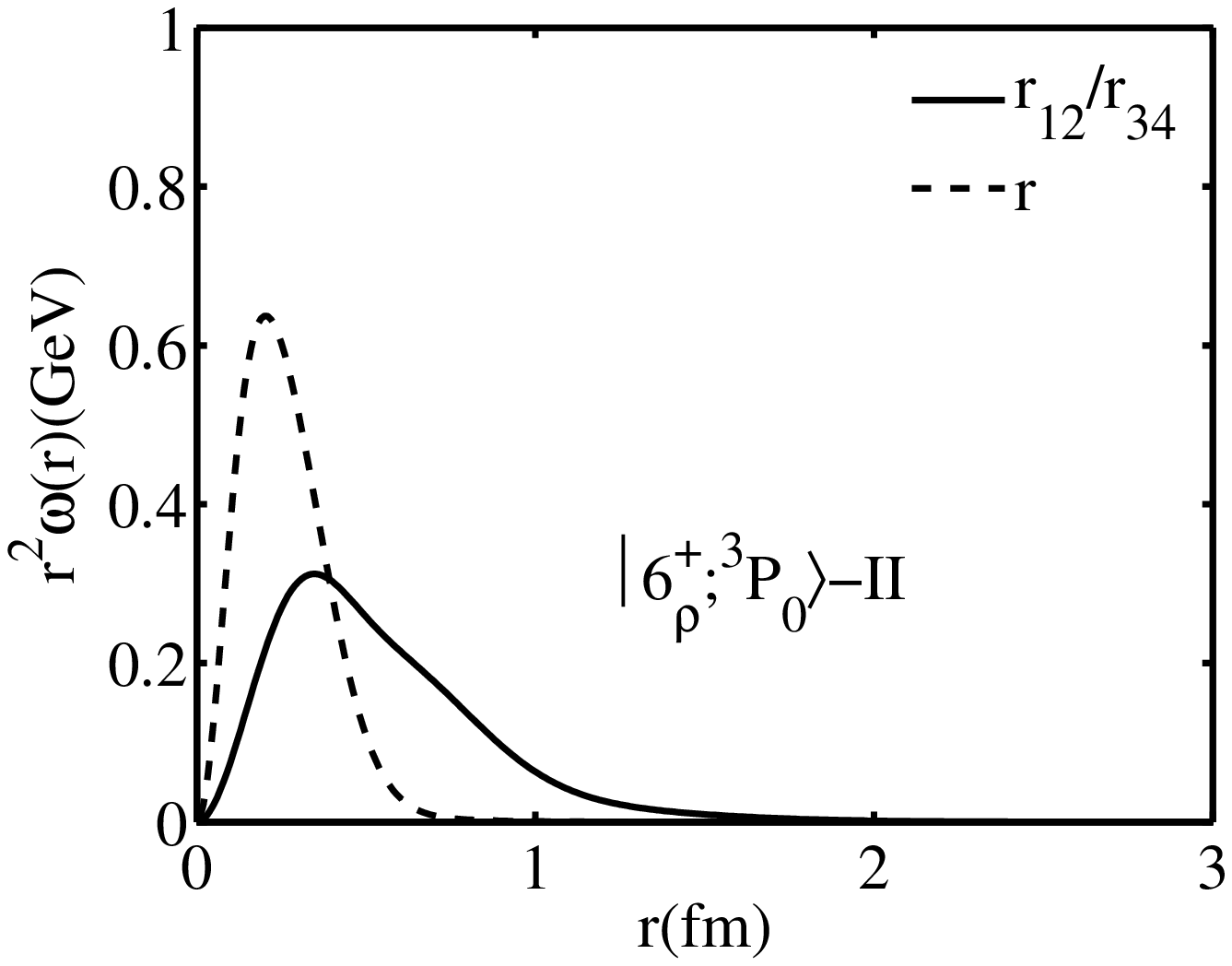}
  \end{minipage}
  }
  \caption{Radial density distributions of the ground state(a) and the first radially excited state(b) of $0^{-+}$($|6^{+}_{\rho};^{3}P_{0}\rangle$)}
\end{figure}

It is shown in Tables V-VI that the masses of the lowest physical states are pulled down by the mixing effect, while the highest states are raised up. The percentage of different color configurations(denoted as basis) of the physical states are listed in the fifth column in Tables VII-VIII. One can see that some physical states have strong mixing effect, while some states are dominated by one component. For example, there are 37.6$\%$ $|3^{+};^{1}S_{0}\rangle$ and 62.4$\%$ $|6^{+};^{1}S_{0}\rangle$ components in the tetraquark state $0^{++}$(6450), while $0^{-+}$(6796) is dominated by 96.6$\%$ $|3^{+}_{\lambda};^{3}P_{0}\rangle$ component.

Up to now, there have been many interpretations about the fully charmed tetraquark states such as the diquark-antidiquark configuration, the meson-meson configuration, molecule states, even mixtures of either two of them, and the coupled-channel effects. Commonly, the $\overline{3}_{c}\otimes3_{c}$ and $6_{c}\otimes\overline{6}_{c}$ configurations are called the diquark-antidiquark configuration, which is illustrated in Fig. 1(a). The representations of Figs. 1(b) and (c) are called the meson-meson configuration, which is given by the product of two color-singlets ($1_{c}\otimes1_{c}$) and the
product of two color-octets ($8_{c}\otimes8_{c}$). As for the $\mathrm{cc}\overline{\mathrm{c}}\overline{\mathrm{c}}$ system, the configurations of Figs. 1(b) and (c) are equivalent to each other. The color relations between the diquark-antidiquark and meson-meson configurations can be expressed as follows,
\begin{eqnarray}
|11\rangle =&\sqrt{\frac{1}{3}}|\overline{3}3\rangle+\sqrt{\frac{2}{3}}|6\overline{6}\rangle \\
|88\rangle =&-\sqrt{\frac{2}{3}}|\overline{3}3\rangle+\sqrt{\frac{1}{3}}|6\overline{6}\rangle
\end{eqnarray}

To investigate the inner structure of the fully charmed tetraquark, we also obtain its proportions in the meson-meson configuration and the r.m.s radii of these states, which are shown in the last four columns of Tables VII-VIII. For $0^{++}$(6450) as an example, it is composed of 37.6$\%$ $|3^{+};^{1}S_{0}\rangle$($\overline{3}_{c}\otimes3_{c}$) and 62.4$\%$ $|6^{+};^{1}S_{0}\rangle$($6_{c}\otimes\overline{6}_{c}$) components. This state contain 54.1$\%$ $1_{c}\otimes1_{c}$ configuration and 45.9$\%$ $8_{c}\otimes8_{c}$ ones. In addition, predicted masses of other collaborations for the ground states of the $\mathrm{cc}\overline{\mathrm{c}}\overline{\mathrm{c}}$ system are listed in Table III. The $8_{c}\otimes8_{c}$ configurations in the table are studied in different QCD sum rules\cite{1S1,1S2} and a lattice-QCD inspired quark model\cite{Tc422}. In Ref.\cite{Tc422}, the author analyzed a complete set of four-body configurations, including diquark-antidiquark, meson-meson, K-type configurations and their couplings. Here, we only list the result of $8_{c}\otimes8_{c}$(meson-meson) configuration. In Ref.\cite{Tc43}, they studied the $\overline{3}_{c}\otimes3_{c}$(diquark-antidiquark) and molecular configurations using the inverse Laplace transform sum rule. Another description for the tetraquark are the butterfly(short for but) and flip-flop configurations adopted in Ref.\cite{1S3}, where their results are larger than those of others. The results in the last column were obtained using a two-body Bethe-Salpeter equation by considering the meson-meson components plus diquark-antidiquark components. It can be seen that our results are roughly compatible with those in Refs.\cite{Tc30,1S1,1S2,Tc422,Tc43}. Our results are consistent well with those of Ref.\cite{Tc30} where they adopted the same method in their calculations except not considering the spin-dependent interactions in the Hamiltonian.
\begin{table}[htp]
\begin{ruledtabular}\caption{Predicted masses of the ground states of tetraquark state in different configurations. All results are in units of GeV}
\begin{tabular}{c c c c c c c c c c c c}
 $J^{PC}$  &  This work  &  \cite{Tc30}& $8_{c}\otimes8_{c}$\cite{1S1} & $8_{c}\otimes8_{c}$\cite{1S2}&$8_{c}\otimes8_{c}$\cite{Tc422}&$\overline{3}_{c}\otimes3_{c}$\cite{Tc43} & Molecule\cite{Tc43} & flip-flop\cite{1S3} & but\cite{1S3} & \cite{1S4}\\ \cline{1-11}
\multirow{2}{*}{$0^{++}$}  & 6.534           & 6.542 & $6.54^{+0.19}_{-0.18}$ &  $6.44^{+0.11}_{-0.11}$  & 6.403&6.411$\pm$0.083 &6.029$\pm$0.198  & 6.850& 6.874 & 5.34\\
\multirow{2}{*}{ }         & 6.450           & 6.435 & $6.36^{+0.16}_{-0.16}$ &  $6.52^{+0.11}_{-0.11}$ & 6.346&6.450$\pm$0.075&6.376$\pm$0.367 & & & \\  \cline{1-11}
                  $1^{+-}$ & 6.517           & 6.515 & $6.47^{+0.18}_{-0.17}$ &                         & 6.325& & & 6.870&6.913 & 6.07\\
                  $2^{++}$ & 6.544           & 6.543 & $6.52^{+0.17}_{-0.17}$ &                         &  6.388& & & 6.913&6.990& \\
\end{tabular}
\end{ruledtabular}
\end{table}

\begin{large}
\textbf{3.2 The newly observed tetraquark states}
\end{large}

In order to exhibit the mass spectrum of the fully charmed tetraquark states more obviously, the results are also displayed in Fig.9. The model predictions for the 1$S$-wave tetraquarks are $0^{++}$(6450), $0^{++}$(6534), $1^{+-}$(6517), and $2^{++}$(6544), which are higher than the $J/\psi J/\psi$ threshold. This indicates that there may not exist bound tetraquark states $\mathrm{cc}\bar{\mathrm{c}}\bar{\mathrm{c}}$ in the scheme of relativistic quark model. From Fig.9, one can see that the mass gaps between the ground states and the first radial excitations are about 300$\sim$400 MeV. This behavior is very similar with that of the doubly charmed baryon spectra. Another important feature about the mass spectrum of the tetraquark states is that the predicted tetraquark states in theory are much richer than the experimental data. One possible explanation about this feature is that some higher resonances have large decay widths and their signals annihilate in the background signals. Even if these resonances are observed in experiments, they may appear as a broad structure in the invariant spectrum. Finally, some states are located very near with each other, e.g. $0^{++}$(6450), $0^{++}$(6534) and $2^{++}$(6544). If the mass splitting of these states is smaller than their decay widths, they will overlap with each other and contribute to a broad structure in the invariant spectrum.
\begin{figure}[h]
\centering
\includegraphics[width=0.9\textwidth]{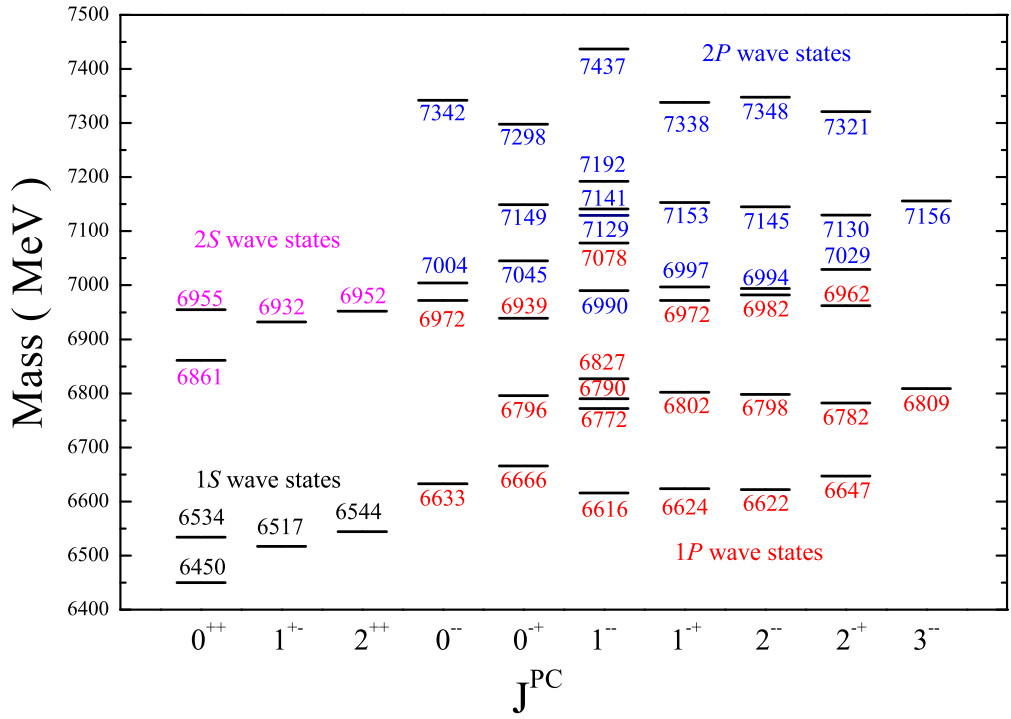}
\caption{Mass spectrum for the $S$-wave and $P$-wave tetraquark states $\mathrm{cc}\overline{\mathrm{cc}}$. The 1$S$, 2$S$, 1$P$ and 2$P$-wave states are displayed in different colors}
\end{figure}
\begin{figure}[h]
\centering
\includegraphics[width=0.9\textwidth]{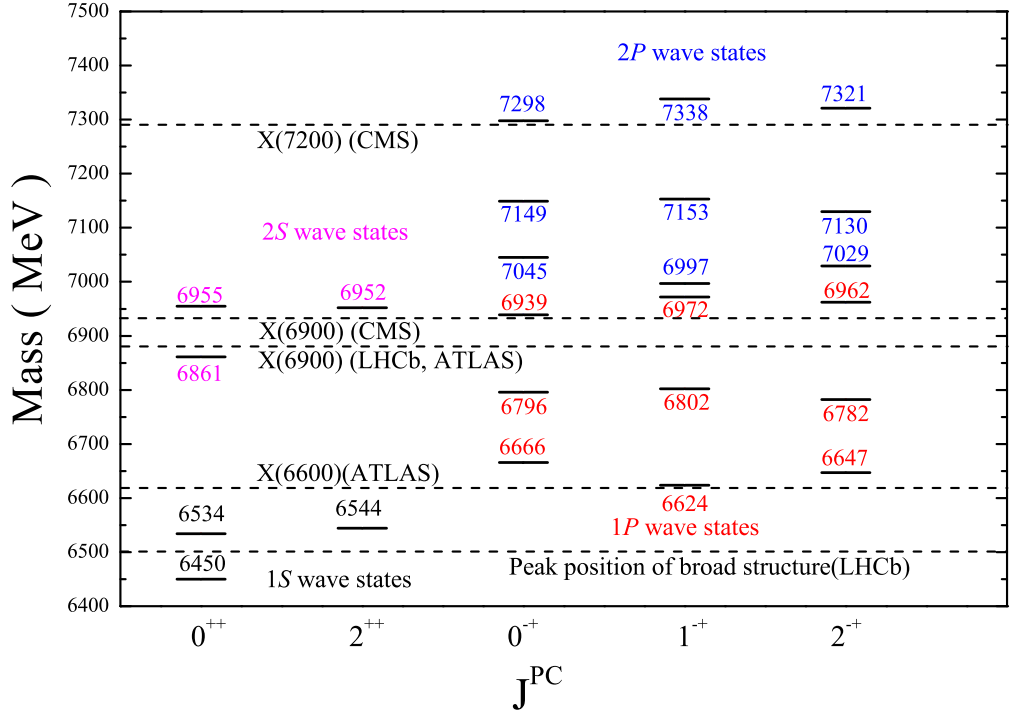}
\caption{Mass spectrum of the full-charmed tetraquark states with positive C-parity.}
\end{figure}

The recently experimental data of the full-charmed tetraquark states were all observed in the $J/\psi J/\psi$ invariant spectrum, which means these new tetraquark states should have positive C-parity. All of the predicted $\mathrm{cc}\bar{\mathrm{c}}\bar{\mathrm{c}}$ states with positive C-parity are shown in Fig.10. The LHCb Collaboration reported a broad structure in the vicinity of 6.5 GeV in 2020. In this energy region, the CMS Collaboration also observed a structure named X(6600) with a measured mass to be $6552\pm10\pm12$ MeV. It is shown in Fig.10 that model predictions for the 1$S$ tetraquark states, $0^{++}$ and $2^{++}$, are located in the range of this structure, thus these 1$S$ states are good candidates for this structure. For the 1$P$-wave tetraquarks, the low-lying $0^{-+}(6666)$, $1^{-+}(6624)$, and $2^{-+}(6647)$ are all located in a structure which was also named as X(6600) by ATLAS Collaboration but with a measured mass to be $6.62\pm0.03^{+0.02}_{-0.01}$ GeV. This structure may be interpreted as a mixture of these low-lying 1$P$ states.

As for the X(6900), the reported results by LHCb, CMS, and ATLAS Collaborations do not consist well with each other. It is shown in Fig.10 that the lowest 2$S$ state $0^{++}$(6861) is consistent with the experimental data $6886\pm11\pm1$ MeV and $6870\pm30^{+60}_{-10}$ MeV. On the other hand, the 2$S$ states $0^{++}$(6955) and $2^{++}$(6952), and a $P$-wave $0^{-+}$(6939) state are compatible with the results $6927\pm9\pm5$ MeV. If X(6900) was treated as a $S$-wave radial excitation, it can decay into the lowest two-charmonium channels $\eta_{c}\eta_{c}$ and $J/\psi J/\psi$ via $S$-wave transition. The decay widths of these decay channels will be broad because of the large phase space. As a $P$-wave tetraquark state, X(6900) can also decay into $J/\psi J/\psi$ channel with a $P$-wave orbital excitation, but its width is kinetically suppressed.

The CMS Collaboration also reported a tetraquark state named as X(7200) with a measuring mass to be $7287\pm19\pm5$. It can be seen in Fig.10 that the high-lying 2$P$ state $0^{-+}$(7298) is located in the range of this structure. This means the $0^{-+}$(7298) is a possible candidate for X(7200) in the framework of relativistic quark model. Considering the decay width of X(7200) plus the uncertainties of the quark model, the $1^{-+}$(7338) and $2^{-+}$(7321) can not be completely excluded. In our present work, we do not include the predicted results of the 3$S$ states. According to the mass gaps between the ground states and the first radial excitations, we can roughly estimate the order of magnitude of 3$S$ states to be 7200$\sim$7300 MeV. This implies the 3$S$ excitations might also be the candidates of X(7200) state.

\begin{Large}
\textbf{4 Conclusions}
\end{Large}

In this work, we have systematically investigate the mass spectra, the r.m.s. radii and the radial density distributions of the $S$- and $P$-wave $\mathrm{cc}\bar{\mathrm{c}}\bar{\mathrm{c}}$ system. The calculation is carried out in the frame work of relativistic quark model, where the Coulomb term, confining potential, tensor potential, contact interaction, and the spin-orbital interaction are all considered. We first analyze the masses, r.m.s. radii, and radial density distributions of different color configurations which are categorized by $\bar{3}_{c}\otimes3_{c}$ and $6_{c}\otimes\bar{6}_{c}$. Then, we obtain the mass spectrum and r.m.s. radii of the physical states by considering the mixing effect.

The results show that the r.m.s. radii of the $\mathrm{cc}\bar{\mathrm{c}}\bar{\mathrm{c}}$ system are less than 1 fm, and the radial density distributions are also in this range. This implies that the full-charmed systems are compact tetraquark states. For the $S$-wave states, the predicted masses are all higher than the thresholds of heavy charmonium pairs, which means no stable bound states exist for $\mathrm{cc}\bar{\mathrm{c}}\bar{\mathrm{c}}$ system. The results also show that the relationship of different color configurations among the mass spectrum is 6$_{\rho}<$3$_{\lambda}<$3$_{\rho}<$6$_{\lambda}$ for $P$-wave full-charmed tetraquarks.

The broad structure in the vicinity of 6.5 GeV reported by LHCb Collaboration and the X(6600) structure with $M_{X(6600)}=6552\pm10\pm12$ MeV may arise from the low-lying 1$S$ tetraquark states $0^{++}$(6450), $0^{++}$(6534) and $2^{++}$(6544). The masses of low-lying 1$P$ $\mathrm{cc}\bar{\mathrm{c}}\bar{\mathrm{c}}$ system with $J^{PC}$=$0^{-+}$, $1^{-+}$, and $2^{-+}$ are predicted to be $6666$, $6624$, and 6647 MeV, respectively. Another tetraquark state also named as X(6600) which were reported with a mass $6.62\pm0.03^{+0.02}_{-0.01}$ MeV by ATLAS Collaboration may be a mixture of these low-lying 1$P$ states. As for the X(6900), the $0^{++}$(6861), $0^{++}$(6955), $2^{++}$(6952), and $0^{-+}$(6939) states are possible candidates for this structure. The definite $J^{PC}$ quantum number for X(6900) can be determined by studying its decay property in the future. Finally, the calculations support assigning X(7200) as a 2$P$ state $0^{-+}$(7298). Certainly, other two 2$P$ states $1^{-+}$(7338), $2^{-+}$(7321) and the 3$S$ states may also be the candidates for this structure.

\begin{Large}
\textbf{Acknowledgments}
\end{Large}

We would like to thank Qi-Fang L\"{u}, Guang-Juan Wang, Xian-Hui Zhong, and Feng-Xiao Liu for their valuable discussions. This project is supported by the Natural Science Foundation of HeBei Province, Grant Number A2018502124.

\begin{table*}[htbp]
\begin{ruledtabular}\caption{Predicted mass (MeV) and the r.m.s.
radius (fm) of different configurations of the S-wave and P-wave tetraquark states. In this table, I and II denote the ground and the first radially excited states, respectively.}
\begin{tabular}{c |c| c| c c c |c c c}
\multirow{2}{*}{L }& \multirow{2}{*}{$J^{PC}$ }& \multirow{2}{*}{Configuration}& \multicolumn{3}{c|}{I} & \multicolumn{3}{c}{II} \\
\multirow{2}{*}{ } & \multirow{2}{*}{ }        & \multirow{2}{*}{}              & M & $\sqrt{\langle r_{12/34}^{2}\rangle}$ & $\sqrt{\langle r^{2}\rangle}$ & M & $\sqrt{\langle r_{12/34}^{2}\rangle}$ & $\sqrt{\langle r^{2}\rangle}$ \\ \hline
\multirow{4}{*}{S-wave  }&\multirow{2}{*}{$0^{++}$}& $|3^{+};^{1}S_{0}\rangle=[[QQ]_{\bar{3}}^{1}[\bar{Q}\bar{Q}]_{3}^{1}]^{0}$ & 6503 & 0.383 &0.309  & 6937 & 0.492 &0.490  \\
\multirow{4}{*}{ }&\multirow{2}{*}{ }& $|6^{+};^{1}S_{0}\rangle=[[QQ]_{6}^{0}[\bar{Q}\bar{Q}]_{\bar{6}}^{0}]^{0}$ & 6482 &0.456 &0.235  & 6879 &0.707 &0.303    \\  \cline{2-9}
\multirow{4}{*}{ } &$1^{+-}$ & $|3^{-};^{3}S_{1}\rangle=[[QQ]_{\bar{3}}^{1}[\bar{Q}\bar{Q}]_{3}^{1}]^{1}$  & 6517 &0.386 &0.310  & 6932 &0.471 &0.528    \\  \cline{2-9}
\multirow{4}{*}{ } &$2^{++}$ & $|3^{+};^{5}S_{2}\rangle=[[QQ]_{\bar{3}}^{1}[\bar{Q}\bar{Q}]_{3}^{1}]^{2}$  & 6544 &0.393 &0.320  & 6952 &0.468 &0.548    \\ \hline
\multirow{20}{*}{P-wave } & \multirow{3}{*}{$0^{-+}$}& $|3^{+}_{\lambda};^{3}P_{0}\rangle=[[[QQ]_{\bar{3}}^{1}[\bar{Q}\bar{Q}]_{3}^{1}]^{1},\lambda]^{0}$ &6796 & 0.412&0.444 & 7146 & 0.457& 0.699 \\
\multirow{20}{*}{ } & \multirow{3}{*}{}& $|3^{+}_{\rho};^{3}P_{0}\rangle=\frac{1}{\sqrt{2}}$($[[[QQ]_{\bar{3}}^{0},\rho]_{\bar{3}}^{1}[\bar{Q}\bar{Q}]_{3}^{1}]^{0}$+c.c.) & 6862 & 0.500 &0.338  & 7232 & 0.573 &0.540  \\
\multirow{20}{*}{ } & \multirow{3}{*}{}& $|6^{+}_{\rho};^{3}P_{0}\rangle=\frac{1}{\sqrt{2}}$($[[[QQ]_{6}^{1},\rho]_{6}^{0}[\bar{Q}\bar{Q}]_{\bar{6}}^{0}]^{0}$+c.c.) & 6743 &0.581 &0.264  & 7114 &0.730 &0.290    \\ \cline{2-9}
\multirow{20}{*}{ } & \multirow{3}{*}{$1^{-+}$}& $|3^{+}_{\lambda};^{3}P_{1}\rangle=|3^{+}_{\lambda};^{3}P_{0}\rangle=[[[QQ]_{\bar{3}}^{1}[\bar{Q}\bar{Q}]_{3}^{1}]^{1},\lambda]^{1}$ &6796 & 0.414&0.444 & 7146 & 0.457& 0.699 \\
\multirow{20}{*}{ } & \multirow{3}{*}{}& $|3^{+}_{\rho};^{3}P_{1}\rangle=\frac{1}{\sqrt{2}}$($[[[QQ]_{\bar{3}}^{0},\rho]_{\bar{3}}^{1}[\bar{Q}\bar{Q}]_{3}^{1}]^{1}$+c.c.) & 6862 & 0.500 & 0.338& 7232& 0.573& 0.540 \\
\multirow{20}{*}{ } & \multirow{3}{*}{}& $|6^{+}_{\rho};^{3}P_{1}\rangle=\frac{1}{\sqrt{2}}$($[[[QQ]_{6}^{1},\rho]_{6}^{1}[\bar{Q}\bar{Q}]_{\bar{6}}^{0}]^{1}$+c.c.) & 6740 & 0.578& 0.263 &7110 &0.727 & 0.289 \\   \cline{2-9}
\multirow{20}{*}{ } & \multirow{3}{*}{$2^{-+}$}& $|3^{+}_{\lambda};^{3}P_{2}\rangle=[[[QQ]_{\bar{3}}^{1}[\bar{Q}\bar{Q}]_{3}^{1}]^{1},\lambda]^{2}$ &6797 & 0.415&0.444 & 7147 & 0.457& 0.700 \\
\multirow{20}{*}{ } & \multirow{3}{*}{}& $|3^{+}_{\rho};^{3}P_{2}\rangle=\frac{1}{\sqrt{2}}$($[[[QQ]_{\bar{3}}^{0},\rho]_{\bar{3}}^{1}[\bar{Q}\bar{Q}]_{3}^{1}]^{2}$+c.c.) & 6862 & 0.500 & 0.338& 7232& 0.573& 0.540 \\
\multirow{20}{*}{ } & \multirow{3}{*}{}& $|6^{+}_{\rho};^{3}P_{2}\rangle=\frac{1}{\sqrt{2}}$($[[[QQ]_{6}^{1},\rho]_{6}^{2}[\bar{Q}\bar{Q}]_{\bar{6}}^{0}]^{2}$+c.c.) & 6732 & 0.573& 0.262 &7101 &0.719 & 0.288 \\  \cline{2-9}
\multirow{20}{*}{ } &\multirow{2}{*}{$0^{--}$}& $|3^{-}_{\rho};^{3}P_{0}\rangle=\frac{1}{\sqrt{2}}$($[[[QQ]_{\bar{3}}^{0},\rho]_{\bar{3}}^{1}[\bar{Q}\bar{Q}]_{3}^{1}]^{0}$-c.c.) & 6862 & 0.500 &0.338  & 7232 & 0.573 &0.540  \\
\multirow{20}{*}{ } &\multirow{2}{*}{ }& $|6^{-}_{\rho};^{3}P_{0}\rangle=\frac{1}{\sqrt{2}}$($[[[QQ]_{6}^{1},\rho]_{6}^{0}[\bar{Q}\bar{Q}]_{\bar{6}}^{0}]^{0}$-c.c.) & 6743 &0.581 &0.264  & 7114 &0.730 &0.290    \\  \cline{2-9}
\multirow{20}{*}{ } & \multirow{5}{*}{$1^{--}$}& $|3^{-}_{\lambda};^{1}P_{1}\rangle=[[[QQ]_{\bar{3}}^{1}[\bar{Q}\bar{Q}]_{3}^{1}]^{0},\lambda]^{1}$ &6791 & 0.411&0.440 & 7142 & 0.458& 0.697 \\
\multirow{20}{*}{ } & \multirow{5}{*}{}& $|3^{-}_{\rho};^{3}P_{1}\rangle=\frac{1}{\sqrt{2}}$($[[[QQ]_{\bar{3}}^{0},\rho]_{\bar{3}}^{1}[\bar{Q}\bar{Q}]_{3}^{1}]^{1}$-c.c.) & 6862 & 0.500 & 0.338& 7232& 0.573& 0.540 \\
\multirow{20}{*}{ } & \multirow{5}{*}{}& $|6^{-}_{\lambda};^{1}P_{1}\rangle=[[[QQ]_{6}^{0}[\bar{Q}\bar{Q}]_{\bar{6}}^{0}]^{0},\lambda]^{1}$ &6883 & 0.529&0.343 & 7249 & 0.715& 0.374 \\
\multirow{20}{*}{ } & \multirow{5}{*}{}& $|6^{-}_{\rho};^{3}P_{1}\rangle=\frac{1}{\sqrt{2}}$($[[[QQ]_{6}^{1},\rho]_{6}^{1}[\bar{Q}\bar{Q}]_{\bar{6}}^{0}]^{1}$-c.c.) & 6740 & 0.578& 0.263 &7110 &0.727 & 0.289 \\
\multirow{20}{*}{ } & \multirow{5}{*}{}& $|3^{-}_{\lambda};^{5}P_{1}\rangle=[[[QQ]_{\bar{3}}^{1}[\bar{Q}\bar{Q}]_{3}^{1}]^{2},\lambda]^{1}$ &6808 & 0.417&0.451 & 7155 & 0.456& 0.704 \\ \cline{2-9}
\multirow{20}{*}{ } & \multirow{3}{*}{$2^{--}$}& $|3^{-}_{\lambda};^{5}P_{2}\rangle=[[[QQ]_{\bar{3}}^{1}[\bar{Q}\bar{Q}]_{3}^{1}]^{2},\lambda]^{2}$ &6808 & 0.417&0.451 & 7155 & 0.456& 0.704 \\
\multirow{20}{*}{ } & \multirow{3}{*}{}& $|3^{-}_{\rho};^{3}P_{2}\rangle=\frac{1}{\sqrt{2}}$($[[[QQ]_{\bar{3}}^{0},\rho]_{\bar{3}}^{1}[\bar{Q}\bar{Q}]_{3}^{1}]^{2}$-c.c.) & 6862 & 0.500 & 0.338& 7232& 0.573& 0.540 \\
\multirow{20}{*}{ } & \multirow{3}{*}{}& $|6^{-}_{\rho};^{3}P_{2}\rangle=\frac{1}{\sqrt{2}}$($[[[QQ]_{6}^{1},\rho]_{6}^{2}[\bar{Q}\bar{Q}]_{\bar{6}}^{0}]^{2}$-c.c.) & 6732 & 0.573& 0.262 &7101 &0.719 & 0.288 \\  \cline{2-9}
\multirow{20}{*}{ } & $3^{--}$ & $|3^{-}_{\lambda};^{5}P_{3}\rangle=[[[QQ]_{\bar{3}}^{1}[\bar{Q}\bar{Q}]_{3}^{1}]^{2},\lambda]^{3}$ &6809 & 0.417&0.451 & 7156 & 0.456& 0.704 \\
\end{tabular}
\end{ruledtabular}
\end{table*}
\begin{table}[htp]
\begin{ruledtabular}\caption{The mass matrix (MeV), the eigenvalue (MeV) and the eigenvector for the ground state by diagonalizing the mass matrix.}
\begin{tabular}{c |c| c |c c c  }
\multirow{2}{*}{L }& \multirow{2}{*}{$J^{PC}$ }&  \multirow{2}{*}{ } & \multicolumn{3}{c}{I}  \\ \cline{4-6}
\multirow{2}{*}{ } & \multirow{2}{*}{ }        &    \multirow{2}{*}{ }   & $ H$     & Eigenvalue &   Eigenvector \\ \hline
\multirow{4}{*}{S-wave  }&\multirow{2}{*}{$0^{++}$}& $|3^{+};^{1}S_{0}\rangle$ & \multirow{2}{*}{$\begin{pmatrix} 6503 & -41 \\ -41 & 6482 \end{pmatrix}$} & 6534 & \multirow{2}{*}{$\begin{matrix} (-0.790,0.613) \\ (0.613,-0.790) \end{matrix}$}   \\
\multirow{4}{*}{ }&\multirow{2}{*}{ }&           $|6^{+};^{1}S_{0}\rangle$ & \multirow{2}{*}{}      & 6450 & \multirow{2}{*}{}  \\  \cline{2-6}
\multirow{4}{*}{ } &$1^{+-}$ &   $|3^{-};^{3}S_{1}\rangle$& (6517) & 6517 & 1  \\  \cline{2-6}
\multirow{4}{*}{ } &$2^{++}$ &   $|3^{+};^{5}S_{2}\rangle$ &(6544) & 6544 & 1 \\ \cline{1-6}
\multirow{20}{*}{P-wave } & \multirow{3}{*}{$0^{-+}$}& $|3^{+}_{\lambda};^{3}P_{0}\rangle$  &\multirow{3}{*}{$\begin{pmatrix} 6796 & 6 & -24 \\ 6&6862 & -77 \\-24 & -77& 6743 \end{pmatrix}$}  &6939 & \multirow{3}{*}{$\begin{matrix} (-0.124,-0.839,0.530 ) \\ (0.983,-0.176, -0.047) \\(0.133, 0.515, 0.847) \end{matrix}$}   \\
\multirow{20}{*}{ } & \multirow{3}{*}{}& $|3^{+}_{\rho};^{3}P_{0}\rangle$ &  & 6796 &    \\
\multirow{20}{*}{ } & \multirow{3}{*}{}& $|6^{+}_{\rho};^{3}P_{0}\rangle$ &  & 6666 &    \\ \cline{2-6}
\multirow{20}{*}{ } & \multirow{3}{*}{$1^{-+}$}& $|3^{+}_{\lambda};^{3}P_{1}\rangle$ &  \multirow{3}{*}{$\begin{pmatrix} 6796 & -14 & 29\\ -14&6862 & 160 \\ 29& 160& 6740 \end{pmatrix}$}  &6972 &  \multirow{3}{*}{$\begin{matrix} (0.028,0.822,0.569 ) \\ (0.983,-0.127, 0.134) \\(-0.182, -0.556, 0.811) \end{matrix}$} \\
\multirow{20}{*}{ } & \multirow{3}{*}{}& $|3^{+}_{\rho};^{3}P_{1}\rangle$ &  & 6802 &  \\
\multirow{20}{*}{ } & \multirow{3}{*}{}& $|6^{+}_{\rho};^{3}P_{1}\rangle$ &  & 6624 &   \\   \cline{2-6}
\multirow{20}{*}{ } & \multirow{3}{*}{$2^{-+}$}& $|3^{+}_{\lambda};^{3}P_{2}\rangle$&  \multirow{3}{*}{$\begin{pmatrix} 6797 & -36 & 72\\ -36 &6862 & -120 \\ 72& -120& 6732 \end{pmatrix}$}  &6962 &  \multirow{3}{*}{$\begin{matrix} (0.391,-0.760,0.518 ) \\ (0.868,0.492, 0.067) \\(-0.307, 0.423, 0.852) \end{matrix}$}\\
\multirow{20}{*}{ } & \multirow{3}{*}{}&  $|3^{+}_{\rho};^{3}P_{2}\rangle$&  & 6782 &  \\
\multirow{20}{*}{ } & \multirow{3}{*}{}&  $|6^{+}_{\rho};^{3}P_{2}\rangle$&  & 6647 &   \\  \cline{2-6}
\multirow{20}{*}{ } &\multirow{2}{*}{$0^{--}$}&  $|3^{-}_{\rho};^{3}P_{0}\rangle$ & \multirow{2}{*}{$\begin{pmatrix} 6862 & 159 \\ 159 & 6743 \end{pmatrix}$} & 6972 &   \multirow{2}{*}{$\begin{matrix} (-0.821,-0.570) \\ (0.570,-0.821) \end{matrix}$}   \\
\multirow{20}{*}{ } &\multirow{2}{*}{ }&  $|6^{-}_{\rho};^{3}P_{0}\rangle$ &  & 6633 &     \\  \cline{2-6}
\multirow{20}{*}{ } & \multirow{5}{*}{$1^{--}$}& $|3^{-}_{\lambda};^{1}P_{1}\rangle$ & \multirow{5}{*}{$\begin{pmatrix} 6791 & -24 & -19 & -49 & -0.3\\ -24 & 6862& 88& 160 & 27 \\ -19 & 88 & 6883 & 94& 1 \\ -49 & 160 & 94 &6740 & 56 \\ -0.3& 27&1 & 56& 6808 \\ \end{pmatrix}$} &7078 &    \multirow{5}{*}{$\begin{matrix} (0.174 , -0.630 , -0.542 , -0.501 , -0.169) \\ (0.039 , -0.194, 0.626, -0.173 , -0.734) \\ (0.861 , -0.084,0.331, -0.077 , 0.369) \\ (-0.449 , -0.585, 0.433, -0.062 , 0.515) \\ (0.157 , -0.466, -0.131, 0.842 , -0.179) \\ \end{matrix}$} \\
\multirow{20}{*}{ } & \multirow{5}{*}{}& $|3^{-}_{\rho};^{3}P_{1}\rangle$ &  & 6827 &  \\
\multirow{20}{*}{ } & \multirow{5}{*}{}&  $|6^{-}_{\lambda};^{1}P_{1}\rangle$ & &6790 &   \\
\multirow{20}{*}{ } & \multirow{5}{*}{}& $|6^{-}_{\rho};^{3}P_{1}\rangle$& & 6772 &   \\
\multirow{20}{*}{ } & \multirow{5}{*}{}&  $|3^{-}_{\lambda};^{5}P_{1}\rangle$&  &6616 &   \\ \cline{2-6}
\multirow{20}{*}{ } & \multirow{3}{*}{$2^{--}$}& $|3^{-}_{\lambda};^{5}P_{2}\rangle$ & \multirow{3}{*}{$\begin{pmatrix} 6808 & 25 & 50\\ 25&6862 & 158 \\ 50& 158& 6732 \end{pmatrix}$} &6982 &  \multirow{3}{*}{$\begin{matrix} (-0.273,-0.787,-0.553 ) \\ (0.950,-0.311, -0.025) \\(-0.152, -0.532, 0.833) \end{matrix}$}  \\
\multirow{20}{*}{ } & \multirow{3}{*}{}& $|3^{-}_{\rho};^{3}P_{2}\rangle$&  & 6798 &  \\
\multirow{20}{*}{ } & \multirow{3}{*}{}& $|6^{-}_{\rho};^{3}P_{2}\rangle$ & & 6622 &   \\  \cline{2-6}
\multirow{20}{*}{ } & $3^{--}$ & $|3^{-}_{\lambda};^{5}P_{3}\rangle$& (6809)&6809 &  1 \\
\end{tabular}
\end{ruledtabular}
\end{table}
\begin{table*}[htbp]
\begin{ruledtabular}\caption{The mass matrix (MeV), the eigenvalue (MeV) and the eigenvector for the first radial excited state by diagonalizing the mass matrix.}
\begin{tabular}{c |c| c |c c c  }
\multirow{2}{*}{L }       & \multirow{2}{*}{$J^{PC}$ } &  \multirow{2}{*}{ }                                           & \multicolumn{3}{c}{II}  \\ \cline{3-6}
\multirow{2}{*}{ }        & \multirow{2}{*}{ }         &  \multirow{2}{*}{ }                                          &  $H$                 & Eigenvalue &   Eigenvector \\ \hline
\multirow{4}{*}{S-wave}   &\multirow{2}{*}{$0^{++}$}   & $|3^{+};^{1}S_{0}\rangle$ &  \multirow{2}{*}{$\begin{pmatrix} 6937 & -37 \\ -37 & 6879 \end{pmatrix}$}                              &     6955      & \multirow{2}{*}{$\begin{matrix} (-0.899,0.438) \\ (-0.438,-0.899) \end{matrix}$}   \\
\multirow{4}{*}{ }        &\multirow{2}{*}{ }          &           $|6^{+};^{1}S_{0}\rangle$ & \multirow{2}{*}{}                                                                                       &      6861     & \multirow{2}{*}{}  \\  \cline{2-6}
\multirow{4}{*}{ }        &$1^{+-}$ &   $|3^{-};^{3}S_{1}\rangle$                   &  (6932)                                                                                                      & 6932      & 1  \\  \cline{2-6}
\multirow{4}{*}{ }        &$2^{++}$ &   $|3^{+};^{5}S_{2}\rangle$                   & (6952)                                                                                                       & 6952      & 1 \\ \cline{1-6}
\multirow{20}{*}{P-wave } & \multirow{3}{*}{$0^{-+}$}  & $|3^{+}_{\lambda};^{3}P_{0}\rangle$ &  \multirow{3}{*}{$\begin{pmatrix} 7146 & 4 & -27 \\ 4 &7232 & -108 \\ -27 & -108& 7114 \end{pmatrix}$}  &7298       & \multirow{3}{*}{$\begin{matrix} (-0.114,-0.849,0.515 ) \\ (0.971,-0.204, -0.122) \\(-0.208, -0.487, -0.848) \end{matrix}$}  \\
\multirow{20}{*}{ }       & \multirow{3}{*}{}          & $|3^{+}_{\rho};^{3}P_{0}\rangle$&  \multirow{3}{*}{}                                                                                      & 7149      &    \\
\multirow{20}{*}{ }       & \multirow{3}{*}{}          & $|6^{+}_{\rho};^{3}P_{0}\rangle$&  \multirow{3}{*}{}                                                                                      & 7045      &    \\ \cline{2-6}
\multirow{20}{*}{ }       & \multirow{3}{*}{$1^{-+}$}& $|3^{+}_{\lambda};^{3}P_{1}\rangle$  &   \multirow{3}{*}{$\begin{pmatrix} 7146 & -20 & 25\\ -20&7232 & 156 \\ 25& 156& 7110 \end{pmatrix}$}        &7338       & \multirow{3}{*}{$\begin{matrix} (0.013,-0.827,-0.563 ) \\ (0.978,-0.107, 0.180) \\(-0.209, -0.553, 0.807) \end{matrix}$}  \\
\multirow{20}{*}{ }       & \multirow{3}{*}{}& $|3^{+}_{\rho};^{3}P_{1}\rangle$          &   \multirow{3}{*}{}                                                                                     & 7153      &  \\
\multirow{20}{*}{ }       & \multirow{3}{*}{} & $|6^{+}_{\rho};^{3}P_{1}\rangle$         &   \multirow{3}{*}{}                                                                                     & 6997      &   \\   \cline{2-6}
\multirow{20}{*}{ }       & \multirow{3}{*}{$2^{-+}$}& $|3^{+}_{\lambda};^{3}P_{2}\rangle$  &   \multirow{3}{*}{$\begin{pmatrix} 7147 & -44 & 65\\ -44 &7232 & -107 \\ 65& -107& 7101 \end{pmatrix}$} &7321       & \multirow{3}{*}{$\begin{matrix} (0.382,-0.782,0.493 ) \\ (0.863,0.493, 0.113) \\(-0.332, 0.382, 0.863) \end{matrix}$}  \\
\multirow{20}{*}{ }       & \multirow{3}{*}{}  &  $|3^{+}_{\rho};^{3}P_{2}\rangle$        &    \multirow{3}{*}{}                                                                                    & 7130      &   \\
\multirow{20}{*}{ }       & \multirow{3}{*}{} &  $|6^{+}_{\rho};^{3}P_{2}\rangle$         &    \multirow{3}{*}{}                                                                                    & 7029      &   \\  \cline{2-6}
\multirow{20}{*}{ }       &\multirow{2}{*}{$0^{--}$} &  $|3^{-}_{\rho};^{3}P_{0}\rangle$  &   \multirow{2}{*}{$\begin{pmatrix} 7232 & 158 \\ 158 & 7114 \end{pmatrix}$}                             & 7342      &   \multirow{2}{*}{$\begin{matrix} (-0.822,-0.570) \\ (0.570,-0.822) \end{matrix}$}  \\
\multirow{20}{*}{ }       &\multirow{2}{*}{ }   &  $|6^{-}_{\rho};^{3}P_{0}\rangle$       &    \multirow{2}{*}{}                                                                                    & 7004 &     \\  \cline{2-6}
\multirow{20}{*}{ }       & \multirow{5}{*}{$1^{--}$} & $|3^{-}_{\lambda};^{1}P_{1}\rangle$  &  \multirow{5}{*}{$\begin{pmatrix} 7142 & -28 & -25 & -46 & -0.2\\ -28 & 7232& 80& 158 & 31 \\ -25 & 80 & 7249 & 86& 1 \\ -46 & 158 & 86 &7110 & 53 \\ -0.2& 31&1 & 53& 7155 \\ \end{pmatrix}$} &7437 & \multirow{5}{*}{$\begin{matrix} (0.184 , -0.641 , -0.526 , -0.501 , -0.166) \\ (-0.012 , -0.330, 0.750, -0.192 , -0.539) \\ (0.709 , -0.185,0.338, -0.054 , 0.588) \\ (-0.664 , -0.474, 0.175, -0.004 , 0.551) \\ (0.148 , -0.470, -0.119, 0.842 , -0.181) \\ \end{matrix}$}  \\
\multirow{20}{*}{ }       & \multirow{5}{*}{}  & $|3^{-}_{\rho};^{3}P_{1}\rangle$        &  \multirow{5}{*}{}                                                                                      & 7192      &  \\
\multirow{20}{*}{ }       & \multirow{5}{*}{} &  $|6^{-}_{\lambda};^{1}P_{1}\rangle$         &   \multirow{5}{*}{}                                                                                     &7141       &   \\
\multirow{20}{*}{ }       & \multirow{5}{*}{} & $|6^{-}_{\rho};^{3}P_{1}\rangle$        &   \multirow{5}{*}{}                                                                                     & 7129      &   \\
\multirow{20}{*}{ }       & \multirow{5}{*}{} &  $|3^{-}_{\lambda};^{5}P_{1}\rangle$         &   \multirow{5}{*}{}                                                                                     &6990       &   \\ \cline{2-6}
\multirow{20}{*}{ }       & \multirow{3}{*}{$2^{--}$} & $|3^{-}_{\lambda};^{5}P_{2}\rangle$ &  \multirow{3}{*}{$\begin{pmatrix} 7155 & 29 & 48\\ 29&7232 & 155 \\ 48& 155& 7101 \end{pmatrix}$}       &7348       &  \multirow{3}{*}{$\begin{matrix} (0.382,-0.782,0.493 ) \\ (0.863,0.493, 0.113) \\(-0.332, 0.382, 0.863) \end{matrix}$} \\
\multirow{20}{*}{ }       & \multirow{3}{*}{}  & $|3^{-}_{\rho};^{3}P_{2}\rangle$        &  \multirow{3}{*}{}                                                                                      & 7145      &  \\
\multirow{20}{*}{ }       & \multirow{3}{*}{}  & $|6^{-}_{\rho};^{3}P_{2}\rangle$        &  \multirow{3}{*}{}                                                                                      & 6994      &   \\  \cline{2-6}
\multirow{20}{*}{ }       & $3^{--}$  & $|3^{-}_{\lambda};^{5}P_{3}\rangle$                 &  (7156)                                                                                                      &7156       &  1 \\
\end{tabular}
\end{ruledtabular}
\end{table*}
\begin{table*}[htbp]
\begin{ruledtabular}\caption{The components of different color configurations(denoted as basis) and the r.m.s. radius(fm) of the ground tetraquark states.}
\begin{tabular}{c |c| c| c c |c c| c c c c}
\multirow{2}{*}{L }       & \multirow{2}{*}{$J^{PC}$ } &  \multirow{2}{*}{Mass}&\multicolumn{4}{c}{I}  \\ \cline{4-11}
\multirow{2}{*}{ }        & \multirow{2}{*}{ }         &     \multirow{2}{*}{}       & Basis &   Components($\%$) &$r_{12}/r_{34}$ &  $r$ & $1_{c}\otimes1_{c}$($\%$)& $8_{c}\otimes8_{c}$($\%$) & $r_{13}/r_{24}$ &  $r^{\prime}$ \\ \hline
\multirow{4}{*}{S-wave  }&\multirow{2}{*}{$0^{++}$}&6534 &$|3^{+};^{1}S_{0}\rangle$    &   \multirow{2}{*}{$\begin{matrix} (62.4,37.6) \\ (37.6,62.4) \end{matrix}$}  & 0.412& 0.283 & 45.9 & 54.1  &0.405 & 0.297\\
\multirow{4}{*}{ }&\multirow{2}{*}{ }& 6450 &$|6^{+};^{1}S_{0}\rangle$                  &   \multirow{2}{*}{} & 0.430&0.265 &54.1 &45.9 & 0.405 & 0.298\\  \cline{2-11}
\multirow{4}{*}{ } &$1^{+-}$ & 6517 &$|3^{-};^{3}S_{1}\rangle$                           &    100 & 0.386& 0.310& 33.3& 66.7 &0.413&0.273\\ \cline{2-11}
\multirow{4}{*}{ } &$2^{++}$ & 6544 &$|3^{+};^{5}S_{2}\rangle$                           &    100 &0.393 & 0.320 & 33.3& 66.7&0.424&0.278\\ \cline{1-11}
\multirow{20}{*}{P-wave } & \multirow{3}{*}{$0^{-+}$}& 6939 &$|3^{+}_{\lambda};^{3}P_{0}\rangle$                  &  \multirow{3}{*}{$\begin{matrix} (1.5,70.4,28.1 ) \\ (96.6,3.2, 0.2) \\(1.8, 26.5, 71.7) \end{matrix}$}  & 0.523 & 0.321 & 42.7& 57.3& 0.489&0.381\\
\multirow{20}{*}{ } & \multirow{3}{*}{}& 6796 &$|3^{+}_{\rho};^{3}P_{0}\rangle$        &  &0.416 & 0.441& 33.4&66.6 &0.512 &0.351\\
\multirow{20}{*}{ } & \multirow{3}{*}{}&  6666 &$|6^{+}_{\rho};^{3}P_{0}\rangle$               &     &0.558  &  0.289 &57.2&42.8 & 0.490& 0.383\\ \cline{2-11}
\multirow{20}{*}{ } & \multirow{3}{*}{$1^{-+}$}& 6972 &$|3^{+}_{\lambda};^{3}P_{1}\rangle$                       &  \multirow{3}{*}{$\begin{matrix} (0.1,67.5,32.4 ) \\ (96.6,1.6, 1.8) \\(3.3, 30.9, 65.8) \end{matrix}$} & 0.493& 0.39& 33.5& 66.5 & 0.490&0.376\\
\multirow{20}{*}{ } & \multirow{3}{*}{}& 6802 &$|3^{+}_{\rho};^{3}P_{1}\rangle$          &   & 0.453& 0.411& 38.7& 61.3&0.509 & 0.354\\
\multirow{20}{*}{ } & \multirow{3}{*}{}& 6624 &$|6^{+}_{\rho};^{3}P_{1}\rangle$                &   &0.555 & 0.300& 61.1& 38.9 &0.507 &0.361\\   \cline{2-11}
\multirow{20}{*}{ } & \multirow{3}{*}{$2^{-+}$}& 6962 &$|3^{+}_{\lambda};^{3}P_{2}\rangle$                      &  \multirow{3}{*}{$\begin{matrix} (15.4,57.8,26.8 ) \\ (75.3,24.2, 0.5) \\(9.4, 17.9, 72.6) \end{matrix}$} & 0.509& 0.339 &42.3&57.7&0.491 &0.373\\
\multirow{20}{*}{ } & \multirow{3}{*}{}& 6782 &$|3^{+}_{\rho};^{3}P_{2}\rangle$         &   & 0.438 & 0.420 &33.5&66.5&0.505 &0.355\\
\multirow{20}{*}{ } & \multirow{3}{*}{}& 6647 &$|6^{+}_{\rho};^{3}P_{2}\rangle$               &   &0.547 & 0.298&57.5&42.4& 0.493&0.373\\  \cline{2-11}
\multirow{20}{*}{ } &\multirow{2}{*}{$0^{--}$}& 6972 &$|3^{-}_{\rho};^{3}P_{0}\rangle$  &    \multirow{2}{*}{$\begin{matrix} (67.4,32.6) \\ (32.6,67.4) \end{matrix}$}  & 0.527& 0.316 & 44.2& 55.8&0.489&0.382\\
\multirow{20}{*}{ } &\multirow{2}{*}{ }& 6633 &$|6^{-}_{\rho};^{3}P_{0}\rangle$               &     & 0.556& 0.290 &55.8 & 44.2&0.489&0.385\\  \cline{2-11}
\multirow{20}{*}{ } & \multirow{5}{*}{$1^{--}$}& 7078 &$|3^{-}_{\lambda};^{1}P_{1}\rangle$                      &    \multirow{5}{*}{$\begin{matrix} (3.0 , 39.7 , 29.3 , 25.1 , 2.8) \\ (0.1 , 3.7, 39.2, 3.0 , 53.9) \\ (74.2 , 0.7, 10.9, 0.6 , 13.6) \\ (20.2 , 34.2, 18.8, 0.4 , 26.5) \\ (2.5 , 21.7, 1.7, 70.9 , 3.2)  \end{matrix}$}& 0.525 & 0.334& 57.0&43.0&0.498&0.366\\
\multirow{20}{*}{ } & \multirow{5}{*}{}& 6827 &$|3^{-}_{\rho};^{3}P_{1}\rangle$         &   & 0.480& 0.380 &40.5&59.5&0.502&0.364\\
\multirow{20}{*}{ } & \multirow{5}{*}{}& 6790 &$|6^{-}_{\lambda};^{1}P_{1}\rangle$                                &  & 0.419 &0.440 & 33.7& 66.3&0.520&0.334\\
\multirow{20}{*}{ } & \multirow{5}{*}{}& 6772 &$|6^{-}_{\rho};^{3}P_{1}\rangle$              &   & 0.471& 0.387& 41.1&58.9&0.510&0.346\\
\multirow{20}{*}{ } & \multirow{5}{*}{}& 6616 &$|3^{-}_{\lambda};^{5}P_{1}\rangle$                               &  & 0.548&0.310 & 61.3&38.7&0.510&0.358\\ \cline{2-11}
\multirow{20}{*}{ } & \multirow{3}{*}{$2^{--}$}& 6982 &$|3^{-}_{\lambda};^{5}P_{2}\rangle$                       &  \multirow{3}{*}{$\begin{matrix} (7.5,61.9,30.6 ) \\ (90.3,9.6, 0.1) \\(2.3, 28.3, 69.4) \end{matrix}$} & 0.518& 0.327&43.5&56.5&0.489&0.376\\
\multirow{20}{*}{ } & \multirow{3}{*}{}& 6798 &$|3^{-}_{\rho};^{3}P_{2}\rangle$         &  & 0.426& 0.441& 33.4&66.6&0.512&0.351\\
\multirow{20}{*}{ } & \multirow{3}{*}{}& 6622 &$|6^{-}_{\rho};^{3}P_{2}\rangle$               &    &0.550 & 0.291 &56.5&43.5&0.488&0.380\\  \cline{2-11}
\multirow{20}{*}{ } & $3^{--}$ & 6809 &$|3^{-}_{\lambda};^{5}P_{3}\rangle$                                        &  100 & 0.417&0.451&33.3&66.7&0.539&0.295\\
\end{tabular}
\end{ruledtabular}
\end{table*}
\begin{table*}[htbp]
\begin{ruledtabular}\caption{The components of different color configurations(denoted as basis) and the r.m.s. radius(fm) of the first radial tetraquark states.}
\begin{tabular}{c |c| c| c c| c c c c c c}
\multirow{2}{*}{L }       & \multirow{2}{*}{$J^{PC}$ } &  \multirow{2}{*}{Masses}&\multicolumn{4}{c}{II}  \\ \cline{4-11}
\multirow{2}{*}{ }        & \multirow{2}{*}{ }         &     \multirow{2}{*}{}       & Basis &  Components ($\%$) &$r_{12}/r_{34}$ &  $r$ & $1_{c}\otimes1_{c}$& $8_{c}\otimes8_{c}$ & $r_{13}/r_{24}$ &  $r^{\prime}$ \\ \hline
\multirow{4}{*}{S-wave}   &\multirow{2}{*}{$0^{++}$}   &  6955      &$|3^{+};^{1}S_{0}\rangle$                                     &      \multirow{2}{*}{$\begin{matrix} (80.8,19.2) \\ (19.2,80.8) \end{matrix}$}  & 0.540& 0.460&39.7&60.3&0.672 &0.425\\
\multirow{4}{*}{ }        &\multirow{2}{*}{ }          & 6861     &$|6^{+};^{1}S_{0}\rangle$                                     &       \multirow{2}{*}{}  &  0.671& 0.347&60.3&39.7&0.592&0.436\\  \cline{2-11}
\multirow{4}{*}{ }        &$1^{+-}$                    &  6932      & $|3^{-};^{3}S_{1}\rangle$                                    &  100 & 0.471& 0.528&33.3&66.7&0.624&0.333\\ \cline{2-11}
\multirow{4}{*}{ }        &$2^{++}$                    & 6952      &$|3^{+};^{5}S_{2}\rangle$                                       &  100 & 0.468& 0.548 &33.3&66.7&0.640&0.331\\ \cline{1-11}
\multirow{20}{*}{P-wave } & \multirow{3}{*}{$0^{-+}$}  & 7298       &$ |3^{+}_{\lambda};^{3}P_{0}\rangle$               & \multirow{3}{*}{$\begin{matrix} (1.3,72.1,26.5 ) \\ (94.3,4.2, 1.5) \\(4.3, 23.7, 71.9) \end{matrix}$}  & 0.617& 0.489&42.2&57.8&0.638 &0.460\\
\multirow{20}{*}{ }       & \multirow{3}{*}{}          &  7149      &$|3^{+}_{\rho};^{3}P_{0}\rangle$     &   & 0.468& 0.689 &33.8&66.2& 0.694&0.422\\ 
\multirow{20}{*}{ }       & \multirow{3}{*}{}          &  7045      &$|6^{+}_{\rho};^{3}P_{0}\rangle$           &   &0.686 &  0.388 &57.3&42.7&0.641&0.462\\ \cline{2-11}
\multirow{20}{*}{ }       & \multirow{3}{*}{$1^{-+}$}  &   7338       &$|3^{+}_{\lambda};^{3}P_{1}\rangle$                         & \multirow{3}{*}{$\begin{matrix} (0.1,68.3,31.7 ) \\ (95.6,1.2, 3.2) \\(4.4, 30.5, 765.1) \end{matrix}$}  & 0.567& 0.552 &33.4&66.6&0.643&0.454\\
\multirow{20}{*}{ }       & \multirow{3}{*}{}          & 7153      & $|3^{+}_{\rho};^{3}P_{1}\rangle$  &  & 0.540& 0.617&41.2& 58.9& 0.687&0.427\\
\multirow{20}{*}{ }       & \multirow{3}{*}{}          & 6997      &  $|6^{+}_{\rho};^{3}P_{1}\rangle$      &   & 0.673& 0.423 &58.7&41.3&0.679&0.436\\   \cline{2-11}
\multirow{20}{*}{ }       & \multirow{3}{*}{$2^{-+}$}  &  7321       & $|3^{+}_{\lambda};^{3}P_{2}\rangle$             & \multirow{3}{*}{$\begin{matrix} (14.6,61.1,24.3 ) \\ (74.5,24.3, 1.3) \\(11.0, 14.6, 74.5) \end{matrix}$} &0.597 &  0.519 &41.4&58.6&0.645&0.450\\
\multirow{20}{*}{ }       & \multirow{3}{*}{}          &   7130      & $|3^{+}_{\rho};^{3}P_{2}\rangle$  &   & 0.492& 0.661 &33.8&66.3& 0.681&0.426\\
\multirow{20}{*}{ }       & \multirow{3}{*}{}          &   7029      & $|6^{+}_{\rho};^{3}P_{2}\rangle$       &   &0.675 &  0.398 &58.2&41.8&0.651&0.449\\  \cline{2-11}
\multirow{20}{*}{ }       &\multirow{2}{*}{$0^{--}$}   &   7342      & $|3^{-}_{\rho};^{3}P_{0}\rangle$   &    \multirow{2}{*}{$\begin{matrix} (67.6,32.4) \\ (32.4,67.6) \end{matrix}$} & 0.628&  0.474&44.1&55.9&0.636&0.462\\
\multirow{20}{*}{ }       &\multirow{2}{*}{ }          &   7004 & $|6^{-}_{\rho};^{3}P_{0}\rangle$       &     & 0.683& 0.389 &55.9&44.1&0.633&0.466\\  \cline{2-11}
\multirow{20}{*}{ }       & \multirow{5}{*}{$1^{--}$}  & 7437 & $|3^{-}_{\lambda};^{1}P_{1}\rangle$              & \multirow{5}{*}{$\begin{matrix} (3.4 , 41.1 , 27.7 , 25.1 , 2.8) \\ (0.1 , 10.9, 56.4, 3.7 , 29.0) \\ (50.3, 3.4,11.4, 0.3 , 34.6) \\ (44.1 , 22.5, 3.1, 0 , 30.3) \\ (2.2, 22.1, 1.43, 71.0 , 3.3)  \end{matrix}$}  & 0.676& 0.427 & 36.5&63.5&0.654&0.453\\
\multirow{20}{*}{ }       & \multirow{5}{*}{}          &  7192      & $|3^{-}_{\rho};^{3}P_{1}\rangle$   &   & 0.574& 0.565&47.0&53.0&0.666&0.444 \\
\multirow{20}{*}{ }       & \multirow{5}{*}{}          &   7141       & $|6^{-}_{\lambda};^{1}P_{1}\rangle$                       &  &0.472 & 0.691&53.6& 36.4&0.656&0.411\\
\multirow{20}{*}{ }       & \multirow{5}{*}{}          & 7129      & $|6^{-}_{\rho};^{3}P_{1}\rangle$         &   & 0.564& 0.587 &35.5&64.5&0.689&0.430\\
\multirow{20}{*}{ }       & \multirow{5}{*}{}          &   6990       & $|3^{-}_{\lambda};^{5}P_{1}\rangle$                        &  &0.677 &  0.420 &60.7&39.3&0.683&0.436\\ \cline{2-11}
\multirow{20}{*}{ }       & \multirow{3}{*}{$2^{--}$}  &  7348       & $|3^{-}_{\lambda};^{5}P_{2}\rangle$                 &  \multirow{3}{*}{$\begin{matrix} (14.6,61.2,24.3 ) \\ (74.5,24.3, 1.3) \\(11.0, 14.6, 74.5) \end{matrix}$} & 0.597& 0.521 & 41.5&58.5&0.645&0.450\\
\multirow{20}{*}{ }       & \multirow{3}{*}{}          & 7145      & $|3^{-}_{\rho};^{3}P_{2}\rangle$    &  &0.491 &  0.664 & 33.7&66.3&0.682&0.426\\
\multirow{20}{*}{ }       & \multirow{3}{*}{}          &  6994      & $|6^{-}_{\rho};^{3}P_{2}\rangle$       &    & 0.675& 0.399&58.2 &41.8&0.651&0.449\\  \cline{2-11}
\multirow{20}{*}{ }       & $3^{--}$                   & 7156       & $|3^{-}_{\lambda};^{5}P_{3}\rangle$                     & 100 & 0.456&  0.704& 33.3&66.7&0.774&0.322\\
\end{tabular}
\end{ruledtabular}
\end{table*}
\end{document}